%% file: main.tex
\documentclass[letterpaper,twocolumn,10pt]{article}
\usepackage{usenix}

\usepackage{tikz}

\usepackage{filecontents}

\usepackage{subcaption}

\usepackage{booktabs}
\usepackage{multirow}
\usepackage{graphicx}
\usepackage{array}
\usepackage{pifont}

\usepackage[ruled,vlined]{algorithm2e}

\usepackage{amsmath,amssymb}

\usepackage{xspace}
\usepackage{textcomp}
\newcommand{\blephasyr}{\textit{BlePhasyr}}

\newcommand{\cmark}{\ding{51}}  

\newcommand{\tightvspace}{\vspace{-0.5cm}}
\begin{document}

\date{}

\title{\texttt{AirCatch}: Effectively tracing advanced tag-based trackers}


\author{
Abhishek Kumar Mishra\\
Inria\\
\texttt{abhishek.mishra@inria.fr}
\and
Swadeep\\
Northeastern University\\
\texttt{lnu.swa@northeastern.edu}
\and
Guevara Noubir\\
Northeastern University\\
\texttt{g.noubir@northeastern.edu}
\and
Mathieu Cunche\\
University of Lyon / Inria\\
\texttt{mathieu.cunche@inria.fr}
}

\maketitle
\begin{abstract}
Tag-based tracking ecosystems help users locate lost items, but can be leveraged for unwanted tracking and stalking. Existing protocol-driven defenses and prior academic solutions largely assume stable identifiers or predictable beaconing. However, identifier-based defenses fundamentally break down against \emph{advanced} rogue trackers that aggressively rotate identifiers. We present \texttt{AirCatch}, a passive detection system that exploits a physical-layer constraint: while logical identifiers can change arbitrarily fast, the transmitter’s analog imprint remains stable and reappears as a \emph{compact} and \emph{persistently occupied} region in Carrier Frequency Offset (CFO) feature space.  \texttt{AirCatch} advances the state of the art along three axes: (i) a \emph{novel, modulation-aware} CFO fingerprint that augments packet-level CFO with content-independent CFO components that amplify device distinctiveness; (ii) a \emph{new tracking detection algorithm} based on high core density and persistence, that is robust to contamination and evasion through per-identifier segmentation; and (iii) an \emph{ultra-low-cost} receiver, a $\approx$\$10 BLE SDR named \blephasyr, built from commodity components, that makes RF-fingerprinting based detection practical in resource-constrained deployments. We evaluate \texttt{AirCatch} across Apple, Google, Tile, and Samsung tag families in multi-hour captures, systematically stress-test evasion using a scenario generator over a grid of transmission and rotation periods, and validate in diverse real-world mobility traces (home/office commutes, public transport, car, and airport travel) while sweeping background tag density. Across these stress tests, \texttt{AirCatch} achieves \emph{no false positives} and \emph{early detection} over a wide range of adversarial configurations and environments, degrading gracefully only in extreme low-rate regimes that also reduce attacker utility. 
\end{abstract}

\input{Sections/Introduction}
\input{Sections/Related_works}
\input{Sections/ThreatModel}
\input{Sections/Designing_Effective_Fingerprints}
\input{Sections/Tracker_Detection_Module}
\input{Sections/Designing_an_Advanced_Tracker}
\input{Sections/Experiments}
\input{Sections/Anti_stalking_tools_testing}
\input{Sections/Evaluation}
\input{Sections/Results}
\input{Sections/Discussion}

\section{Conclusion}
With the proliferation of tag-based ecosystems, advanced trackers increasingly get the opportunity of evading current OS-level protections by rotating identifiers and duty-cycling transmissions. We introduce \texttt{AirCatch}, which shows that analog-layer invariants can restore reliable tracking detection even when the digital surface is adversarially unstable. We present the first end-to-end defense that extracts stable CFO fingerprints from commodity BLE advertisements, maps them into robust embeddings, and detects active evasion by flagging persistent, abnormally compact CFO cores that repeatedly accumulate short-lived identifiers across time, shifting detection from brittle identifier logic to a physically grounded signature that is difficult to erase without sacrificing radio functionality. Beyond a single system, \texttt{AirCatch} contributes a practical, reproducible foundation for the community: a low-cost BLE SDR named \blephasyr, a principled tracker algorithm that separates benign tags from advanced trackers, and an easy-to-use anti-tracking application. As tag-based stalking scales, \texttt{AirCatch} advances the state of the art by closing the gap between current assumptions and real adversarial behavior.




\section*{Ethical Considerations}

Our study examines the detectability of commodity BLE trackers and introduces \texttt{AirCatch}, a receiver-side detector operating solely on broadcast BLE advertisements observed in the environment. We collected traffic in four real-world mobility traces and in controlled experiments using our own ESP32-based trackers, strictly observing bystander devices without interaction, connection, targeting, or attempts to identify individuals, and limiting measurements to public or semi-public spaces reflecting ordinary mobility conditions. To minimize privacy risk, we do not retain raw identifiers in any released artifacts: all MAC addresses and ecosystem-specific identifiers (e.g., Samsung PRIVIDs) are keyed-hashed and truncated using per-capture keys that are destroyed after processing, preventing cross-dataset linkage or re-identification. Advertisement payloads and raw IQ samples are used only transiently to compute physical-layer features (e.g., CFO-derived statistics) and are discarded thereafter; downstream analysis retains only derived, non-content features and aggregate statistics, without storing user identities or application-layer data.

\section*{Open Science}
\noindent
We make all artifacts required to evaluate and reproduce the contributions of this paper available through an anonymous, public repository:
\begin{center}
\texttt{https://anonymous.4open.science/r/AirCatch-7043/}
\end{center}
\noindent
The repository is fully self-contained and includes the complete implementation of our end-to-end pipeline, scripts to reproduce the main experiments, and comprehensive documentation describing how to execute each stage. In accordance with the Open Science Policy, the artifact provides a detailed README along with all code necessary for data capture across supported platforms, the detection algorithm, and the accompanying mobile application, enabling independent validation and reuse by the community.

\bibliographystyle{plain}
\bibliography{references}

\include{Sections/appendix}

\end{document}

%% file: Sections/Introduction.tex
\section{Introduction}
\label{sec:intro}

BLE-based item trackers (e.g., Apple AirTag, Samsung SmartTag, Google Find My Device tags, and Tile) have enabled crowd-sourced localization at a global scale, but have also been repeatedly repurposed for unwanted tracking and surveillance ~\cite{ceccio2023sneaky,roth2022airtag,samsung_usenix,google_tracker,tile_georgia_tech}. To mitigate abuse, ecosystems combine (i) protocol privacy mechanisms (rotating identifiers, cryptographic payloads) and (ii) platform/user-facing alerts intended to surface tags that move along with a victim for an extended period~\cite{airguard,Apple_trackers_aanjhan,gerhardt2025airtag}. In practice, these defenses remain uneven and often do not work across ecosystems, and can be slow under real mobility and dense public settings~\cite{gerhardt2025airtag,jang2025tale}. Moreover, they largely depend on \emph{protocol-visible continuity}, MAC/payload correlation across time, which is exactly what an advanced rogue tracker can disrupt by rapidly rotating pseudonyms and duty-cycling transmissions (a capability already achievable with modified tags and open tooling)~\cite{heinrich2021openhaystack,mayberry2023blind}. As a result, stealthy attackers preserve tracking utility while depriving identifier-based detectors of persistent evidence.

We present \texttt{AirCatch}, a passive detector designed for this \emph{advanced} regime. \texttt{AirCatch} exploits a physical-layer constraint: while logical identifiers can change arbitrarily fast, a transmitter’s analog imprint is far harder to shed. In particular, the carrier frequency offset (CFO) induced by oscillator and synthesizer imperfections tends to reappear as a compact region in CFO space for a given device, even as MACs and payload fields churn. \texttt{AirCatch} turns this into an operational signature of unwanted tracking: an adversary that follows a victim induces a \emph{persistently occupied, abnormally compact CFO core} that repeatedly accumulates many short-lived identifiers. To make this robust in heterogeneous, crowded traces, \texttt{AirCatch} (i) introduces a modulation-aware CFO fingerprint that augments packet-level CFO with content-independent CFO components to amplify distinctiveness, (ii) uses contamination-resistant segmentation and per-ecosystem normalization before clustering in CFO space, and (iii) applies a persistence plus core-density based detection algorithm aligned with user-facing anti-tracking requirements. 

Unlike prior CFO-enabled BLE defenses aimed at detecting spoofing/masquerade against stationary devices (e.g., BlueShield and BLEGuard, which modify commodity sniffers such as Ubertooth and rely on features like advertising interval, RSSI, and CFO)~\cite{wu2020blueshield,cai2024securing}, \texttt{AirCatch} targets \emph{co-moving tracker detection under identifier rotation} across real mobility contexts and dense backgrounds, where transient encounters and ecosystem heterogeneity dominate. We also explicitly account for realistic background emitters observed in practice (e.g., AirPods-like behavior) when assessing robustness~\cite{muller2025airtagged}.

We evaluate \texttt{AirCatch} across various tag families with multi-hour captures, systematically stress-test evasion using a scenario generator over a grid of transmission periods (with per-transmission pseudonym rotation), and validate in diverse real mobility traces (home/office commutes, public transport, car travel) plus an airport stress test for false positives in a dense public environment. Across these stress tests, \texttt{AirCatch} cleanly separates benign background traffic from advanced tracking behavior at conservative operating points, achieving zero false positives and reliable detection over a wide range of attacker configurations. It can potentially only degrade gracefully in extremely low-rate regimes that also reduce attacker utility. Our major contributions are fourfold:

\vspace{0.2cm}
\noindent\textbf{1. Modulation-aware RF fingerprints.} We introduce a \emph{holistic}, modulation-aware fingerprint that augments packet-level CFO with content-independent CFO components, yielding a stable and highly discriminative physical-layer signature. 

\vspace{0.2cm}
\noindent\textbf{2. Passive detection of advanced rogue trackers.} We present \texttt{AirCatch}, a passive anti-tracking detector that remains effective under \emph{per-transmission identifier rotation} and \emph{duty-cycled/stealthy} beaconing by leveraging a stable signal: a persistently compact and densely occupied CFO core.

\vspace{0.2cm}
\noindent\textbf{3. Low-cost SDR-based BLE anti-tracking device.} We design and release \blephasyr, a pocket-size, $\approx$\$10 BLE \emph{micro-SDR} that provides CFO-grade physical-layer evidence for passive monitoring, making anti-tracking practical for everyday, resource-constrained deployments.

\vspace{0.2cm}
\noindent\textbf{4. End-to-end system and real-world validation.} We provide a companion Android application that surfaces actionable user alerts and optional technical insights, and we demonstrate robust \texttt{AirCatch} performance via an extensive evaluation spanning multiple tracker ecosystems, adversarial configurations, and real-world mobility environments.





%% file: Sections/Related_works.tex
\section{Related Works}
\label{sec:related}

\paragraph{Commodity BLE tracker ecosystems and protocol behavior.}
A growing body of work reverse-engineers and characterizes crowd-sourced finding ecosystems, including Apple AirTag~\cite{roth2022airtag,martin2019handoff}, Samsung SmartTag~\cite{samsung_usenix}, Google Find My Device~\cite{google_tracker}, and Tile~\cite{tile_georgia_tech}. Beyond protocol structure, recent studies analyze real-world behavior and user-facing safety properties such as alert latency, delivery delays, and cross-ecosystem coverage~\cite{heinrich2024please,gerhardt2025airtag,turk2024stop,jang2025tale}. Prior work also highlights additional ways these ecosystems get repurposed (e.g., leveraging tags for localization)~\cite{hany2024airtags} or turning commodity devices into emitters compatible with offline-finding infrastructures~\cite{any_device_to_airtag}. We complement these efforts with a consolidated protocol-level analysis of identifier rotation and safety-relevant state behavior across ecosystems (\S\ref{sec:tracker_behavior}), distilling the privacy and linkability implications that motivate our threat model. While these efforts clarify the operational landscape and design trade-offs, they do not provide a solution that remains reliable when a tracker \emph{actively} manipulates identifier exposure (fast MAC rotation and stealthy transmission) across heterogeneous environments. 

\tightvspace
\paragraph{Identifier-layer fingerprinting, linkability, and evasion.}
At the MAC layer, BLE includes standardized address randomization, and empirical studies confirm that major mobile platforms implement it effectively~\cite{bluetooth_tracking_becker}. However, linkability can persist through ecosystem-specific design choices and unsynchronized rotations between MAC and payload fields; for instance, prior work (and our observations) reports long-lived identifiers in some tracker families (notably Tile)~\cite{tile_georgia_tech,pace2023every}. More fundamentally, custom emitters can directly manipulate advertisement identifiers and rotation schedules, undermining any detector that assumes continuity in protocol-visible fields~\cite{heinrich2021openhaystack}. As a result, identifier-layer correlation alone is not reliable for anti-tracking against an \emph{active} adversary.

\tightvspace
\paragraph{Anti-stalking defenses.}
Platform defenses and third-party tools attempt to surface co-moving trackers across ecosystems, including cross-ecosystem detection on Android for Apple Find My devices~\cite{airguard} and analyses of Apple’s safety alerts and limitations~\cite{Apple_trackers_aanjhan}. Additional systems propose improvements to detection and localization workflows or measurement of network reporting behavior~\cite{muller2023homescout,muller2025smart,briggs2022ble,despres2023detagtive,ibrahim2023tag}. Broader surveys of covert tracking and intimate partner surveillance emphasize that commodity surveillance devices remain readily accessible and that detection tooling is uneven in practice~\cite{ceccio2023sneaky}. Recent work also explores algorithmic directions for improving anti-abuse detection pipelines~\cite{eldridge2024abuse,mayberry2023blind}. Nonetheless, these approaches predominantly rely on identifier continuity and/or ecosystem cooperation, and thus offer limited guarantees of fast pseudonym rotation and rate manipulation.

\tightvspace
\paragraph{Protocol analyses and physical-layer fingerprinting.}
Security analyses of BLE proximity tracking protocols identify design choices and attack surfaces that influence both privacy and detectability~\cite{liu2025thorough}. Separately, physical-layer fingerprinting leverages hardware impairments such as CFO, IQ imbalance, and transients~\cite{brik2008wireless}. In BLE, physical-layer signatures (including CFO) have been shown to enable device tracking~\cite{cfo_ucsd}, and subsequent work explores CFO-derived features for device discrimination or tracking~\cite{sun2017cv,stoian2025augmenting}. Defenses that obfuscate physical-layer fingerprints (including CFO manipulation) have also been explored for certain chipsets~\cite{nikoofard2022protecting}. However, prior work does not address the core anti-stalking requirement we target: a \emph{deployable, low-cost} solution that produces evidence of \emph{persistent} tracking behavior under adversarial identifier churn.

\tightvspace
\paragraph{Summary.}
Prior work (i) characterizes tracker ecosystems and their safety mechanisms, (ii) evaluates and improves identifier-driven detectors, (iii) demonstrates that motivated adversaries can evade identifier-layer continuity, and (iv) shows that physical-layer signatures can potentially enable tracking. What remains critically missing is a framework designed for detecting the trackers that follow the \emph{advanced} regime: per-transmission identifier rotation and adjustable transmission schedules. \texttt{AirCatch} fills this gap by grounding detection in \emph{persistent physical structure} rather than protocol-visible identifiers, while remaining low-cost. 

%% file: Sections/ThreatModel.tex
\section{Threat Model and Adversary}
\label{sec:threat_model}

We study detection of \emph{rogue BLE trackers} that remain co-located with a victim over time while \emph{actively evading protocol-layer linkability}. The defender is a small commodity passive sniffer that observes BLE advertisements and extracts timestamps $t_k$, advertiser addresses $m_k$, payloads $p_k$, and CFO-derived feature vectors $\mathbf{f}_k$ (Section~\ref{sec:fingerprints}). The detector flags \emph{persistently occupied} and \emph{abnormally compact} regions in CFO-feature space that aggregate evidence across many short-lived identifiers (Section~\ref{sec:detection_module}).

\tightvspace
\paragraph{Goals.}
The primary goal is tracking safety: detect a nearby device that persistently follows the victim, even under identifier churn. Operationally, we target near-zero false alarms in crowded environments. We do not aim to perfectly separate \emph{all} surrounding devices or attribute a flagged device to a specific owner; rather, we detect the signature of tracking as joint CFO-space compactness and persistence. 

\tightvspace
\paragraph{Defender assumptions.}
We assume the adversary's transmissions are observable with non-trivial probability when the tracker is near the victim (a requirement for tracking functionality). We do not assume calibrated receivers or access to cloud identifiers, pairing secrets, or stable BLE addresses; our evaluation spans heterogeneous receivers, and our pipeline uses within-type normalization and robust statistics to mitigate receiver bias and transient outliers.

\tightvspace
\paragraph{Tracker assumptions.}
We assume the modified tracker mirrors Lost Mode behavior. Apple Tags transmit the public key only in Lost Mode, while others set a Lost Mode bit that alters identifier rotation. We consider only devices in Lost Mode.

\subsection{Adversary Model}
\label{subsec:adversary_model}

\paragraph{Objective.}
The adversary's goal is to track the victim by keeping a device co-located for extended time while avoiding detection by rotating identifiers, reducing linkability, and shaping emissions.

\tightvspace
\paragraph{Capabilities.}
We assume firmware-level control typical of a determined attacker: (i) arbitrary advertiser-address rotation and payload-field changes, (ii) duty-cycling and rate shaping over a wide range (seconds to minutes), (iii) emission-pattern shaping to mimic benign traffic, and (iv) operation under mobility and interference (home/office, public transport, car). We assume the adversary knows the detection algorithm and thresholds and may tune behavior accordingly.

\tightvspace
\paragraph{Physical and functional constraints.}
Despite protocol control, the adversary faces two constraints exploited by our design. First, for a fixed transmitter, oscillator and analog impairments induce a CFO signature that is stable relative to identifier churn (Section~\ref{sec:fingerprints}); rotating identifiers does not change the underlying hardware imprint. Second, meaningful tracking requires \emph{persistence}: the tracker must reappear across time and contexts. Extended periods of silence reduces tracking utility; continued functionality forces repeated emissions that accumulate evidence across windows.

\tightvspace
\paragraph{Threats addressed.}
Our detector is designed to withstand: (i) \emph{fast identifier rotation}: segmentation prevents cross-device contamination and rotation increases the number of segments contributed by the same physical emitter, strengthening persistence/density evidence; (ii) \emph{duty-cycling}: we quantify and tune robustness in low-support regimes via windowing and robust embedding features; (iii) \emph{crowded environments}: within-type clustering and persistence thresholds reject transient passer-bys even when locally dense; and (iv) \emph{mobility variation}: robust deviation features (median absolute deviation (MAD) and interquartile range (IQR)) reduce sensitivity to interference bursts while preserving CFO geometry.

\subsection{Limitations and Out-of-Scope Attacks}
\label{subsec:out_of_scope}

\paragraph{Hardware-level CFO obfuscation.}
An adversary equipped with transmitter-side countermeasures that deliberately randomize or compensate CFO can reduce fingerprint stability and may evade CFO-based grouping. We treat this as a stronger attacker class requiring non-trivial hardware or firmware modifications and leave it as future work. 

\tightvspace
\paragraph{Near-total silence and receiver manipulation.}
If a tracker transmits so rarely that it falls below practical capture probability, any passive detector will fail; such operation also limits tracking utility, and our results quantify this trade-off. Receiver jamming and physical compromise are out of scope. 

%% file: Sections/Designing_Effective_Fingerprints.tex
\section{Designing Effective Fingerprints}
\label{sec:fingerprints}

This section describes the fingerprint design used throughout our study. Our goal is to derive \emph{device-distinct} and \emph{time-stable} features from passively observed BLE packets, while remaining robust to channel variation, payload changes, and aggressive identifier rotation. We focus on CFO as a physically grounded proxy for oscillator-level imperfections, and we further decompose CFO by \emph{symbol transition type} to expose additional hardware-dependent structure.

\subsection{Challenges of State-of-the-Art Solutions}
\label{subsec:fingerprint_challenges}

Prior work on radiometric fingerprinting has demonstrated that RF front-end imperfections (e.g., oscillator offset, IQ imbalance, phase noise, PA non-linearities) can form device-distinct signatures under passive observation~\cite{brik2008wireless}. However, applying these techniques to commodity BLE trackers raises several practical challenges:

\tightvspace
\paragraph{Protocol-layer churn and short packets.}
Modern tracking ecosystems rotate identifiers and may randomize advertisement payloads, making fingerprints based on stable identifiers, long preambles, or repeated known sequences brittle when observable structure is short-lived or partially unknown.

\tightvspace
\paragraph{Channel confounding.}
Many fingerprints rely on amplitude- or spectrum-based features (e.g., PSD shape, transient energy, rise/fall times). These can vary significantly with multipath, relative motion, antenna orientation, and receiver front-end filtering, confounding device effects with environment effects. Robust designs should minimize sensitivity to unknown channel gain and phase.

\tightvspace
\paragraph{Receiver dependence and sampling constraints.}
Fine-grained transient/spectral approaches often require high-rate IQ captures, consistent receiver calibration, and strict alignment across packets. In contrast, our setting targets practical deployments: commodity receivers, varying capture quality, and heterogeneous packet contents.

\tightvspace
\paragraph{Adversarial adaptation.}
Recent work explicitly identifies CFO as a privacy-relevant fingerprint in BLE and proposes hardware countermeasures that \emph{actively obfuscate} CFO, indicating both its strength and its adversarial relevance~\cite{nikoofard2022protecting}. Thus, a robust fingerprint must exploit CFO while also extracting additional structure beyond a single scalar offset.

\subsection{CFO as a Proxy}
\label{subsec:cfo_proxy}

CFO mainly stems from transmitter–receiver oscillator mismatch and appears as a constant residual rotation in the received complex baseband. Let the received samples be
\begin{equation}
r[n] \;=\; s[n]\;e^{j(2\pi \Delta f \, n/f_s + \phi_0)} \;+\; w[n],
\label{eq:cfo_model}
\end{equation}
where $s[n]$ is the transmitted baseband waveform, $\Delta f$ is the carrier-frequency offset, $f_s$ is the sampling rate, and $w[n]$ is noise. Under this model, the one-step product
\begin{equation}
z[n] \;=\; r[n]\;r^\ast[n-1]
\end{equation}
has expected phase $\arg(z[n]) \approx 2\pi \Delta f/f_s$ plus modulation/noise terms. Summing these phasors across a packet suppresses zero-mean modulation contributions and yields a low-complexity CFO estimate:
\begin{equation}
\widehat{\Delta f}_{\textsf{packet}}
\;=\;
\frac{f_s}{2\pi}\;
\arg\!\Big(\sum_{n=1}^{N-1} r[n]\;r^\ast[n-1]\Big).
\label{eq:cfo_quick}
\end{equation}
This estimator is attractive for BLE-scale packets: it is $O(N)$, requires no pilots, and is compatible with gated windows (e.g., restricting to structurally reliable sub-regions of a packet). CFO has been used as a stable physical-layer identifier and as a discriminative signal statistic in multiple wireless contexts, including BLE~\cite{sun2017cv,albehadili2024performance}.

\tightvspace
\paragraph{Packet-window CFO.}
Our implementation computes CFO over an \emph{exact IQ slice} corresponding to a decoded packet (bounded by decoder-provided sample indices) and optionally applies gating to reduce out-of-packet samples. On the gated window $x=\{x[n]\}_{n=0}^{N-1}$, we compute CFO as $\widehat{\Delta f}_{\textsf{packet}}$ (Eq.~\ref{eq:cfo_quick}).

\subsection{Transition-Type CFOs as Distinct Features}
\label{subsec:transition_cfos}

A key limitation of a single CFO scalar is that it primarily captures the oscillator mismatch $\Delta f$, which may be similar across devices from the same vendor/batch. To extract additional device-dependent structure without relying on payload semantics, we condition CFO estimation on \emph{bit transition types}. Transition-conditioned features can significantly improve identifiability by capturing imperfections that manifest during symbol transitions.  

In Gaussian Frequency Shift Keying (GFSK), the instantaneous frequency is shaped by a Gaussian filter whose impulse response spans multiple symbol intervals. As a result, the discriminator output at time $n$ reflects not only the current bit $b_i$, but also residual phase and frequency contributions from preceding symbols. For \emph{equal transitions} ($0\!\rightarrow\!0$ and $1\!\rightarrow\!1$), the frequency trajectory remains on the same side of the modulation center, and the filter state reinforces a steady frequency bias dominated by the transmitter’s oscillator offset and static analog impairments. 

In contrast, \emph{jump transitions} ($0\!\rightarrow\!1$ and $1\!\rightarrow\!0$) force a rapid sign change in the frequency deviation, causing transient asymmetries in the discriminator response due to finite filter memory, PLL settling dynamics, and DAC/PA slew limitations. These asymmetries are hardware-dependent and persist across packets, even when protocol-layer identifiers and payloads change. By estimating CFO separately over equal and jump transitions using the same phasor-sum estimator, we isolate these transition-dependent biases, yielding features that are both physically grounded and complementary to packet-level CFO.

\tightvspace
\paragraph{Bit-aligned partitioning.}
Let $b_i\in\{0,1\}$ denote the recovered bit sequence for a packet (excluding preamble in our extraction), and let each bit correspond to $sps$ complex samples. Define the symbol-$i$ sample window as
\begin{equation}
\mathcal{W}_i = \{\,n \mid n \in [i\cdot sps, (i+1)\cdot sps)\,\}.
\end{equation}
For each transition $(b_{i-1}\!\rightarrow b_i)$, we assign the phasor products
$x[n]x^\ast[n-1]$ for $n\in \mathcal{W}_i$ (including the boundary product at the start of $\mathcal{W}_i$) to one of four accumulators:
\begin{equation}
T_{00},\;T_{11},\;T_{10},\;T_{01},
\end{equation}
corresponding to transitions $0\!\rightarrow\!0$, $1\!\rightarrow\!1$, $1\!\rightarrow\!0$, and $0\!\rightarrow\!1$, respectively.

\tightvspace
\paragraph{Transition CFO estimators (phasor-sum form).}
For each transition class $uv\in\{00,11,10,01\}$, we compute a CFO estimate using the \emph{same estimator form} as Eq.~\ref{eq:cfo_quick}, but restricted to the class-specific phasor set:
\begin{equation}
\widehat{\Delta f}_{uv}
\;=\;
\frac{f_s}{2\pi}\;
\arg\!\Big(\sum_{(n\in T_{uv})} x[n]\;x^\ast[n-1]\Big).
\label{eq:cfo_transition}
\end{equation}
This is exactly the computation implemented by our transition CFO module: it sums the complex products for samples belonging to each transition type and converts the resulting phasor angle into Hz. Importantly, using boundary products ensures that the union of all class-specific products recovers the global phasor sum:
\begin{equation}
\sum_{uv}\sum_{n\in T_{uv}} x[n]x^\ast[n-1]
\;=\;
\sum_{n=1}^{N-1} x[n]x^\ast[n-1],
\end{equation}
so that the recombined estimate equals $\widehat{\Delta f}_{\textsf{packet}}$ computed on the same IQ window. This provides an internal consistency check and prevents ``feature drift'' caused by misaligned partitions. 

\subsection{Holistic Fingerprints}
\label{subsec:holistic}

We define a \emph{holistic} fingerprint vector that jointly captures (i) the packet-level CFO (global oscillator offset proxy) and (ii) symbol–conditioned CFOs (symbol-dynamics proxy). 

For each decoded packet with gated complex samples, we extract the following $5$-dimensional fingerprint:
\begin{equation}
\mathbf{f}_{\textsf{IQ}}
=
\big[
\widehat{\Delta f}_{\textsf{packet}},\;
\widehat{\Delta f}_{00},\;
\widehat{\Delta f}_{11},\;
\widehat{\Delta f}_{10},\;
\widehat{\Delta f}_{01}
\big],
\label{eq:holistic_fingerprint_iq}
\end{equation}
where $\widehat{\Delta f}_{\textsf{packet}}$ is the packet-level CFO estimated using the phasor-sum estimator (Eq.~\ref{eq:cfo_quick}), and $\widehat{\Delta f}_{ab}$ denotes the transition-conditioned CFO for bit transitions $a\!\rightarrow\!b$, computed using Eq.~\ref{eq:cfo_transition} over the \emph{same gated IQ window}. 

In settings where full-rate complex IQ is not available (e.g., Ubertooth-class devices), the receiver exposes only a coarse per-byte frequency proxy rather than continuous complex samples. For this case, we engineer an alternative \emph{content-stratified} CFO fingerprint based on Hamming-weight conditioning of the proxy; we describe the construction, rationale, and high effectiveness in Appendix~\S\ref{app:ubertooth}. Intuitively, this stratification recovers discriminative structure that is otherwise averaged out by a single coarse CFO mean, providing a practical fallback when only limited radio statistics are observable. In this paper, our primary focus is on \blephasyr, our ultra-low-cost BLE \emph{micro-SDR} that provides full I/Q for robust CFO and transition-CFO extraction. 

\tightvspace
\paragraph{Novelty.}
Our holistic fingerprint departs from prior CFO-only approaches in two fundamental ways. First, we introduce a \emph{modulation-aware decomposition} of CFO via explicit bit transitions, grounded in the same physical origin: transmitter-specific modulation dynamics shaped by Gaussian filtering and analog front-end imperfections. Second, we treat these conditioned CFOs as \emph{first-class radiometric features}, yielding a compact and interpretable $5$-dimensional representation that is robust to protocol-layer identifier rotation and independent of payload semantics. This directly targets the operational regime of BLE trackers, where identifiers and payloads evolve rapidly, but hardware-induced impairments persist across time and packets~\cite{brik2008wireless,nikoofard2022protecting}. 

%% file: Sections/Tracker_Detection_Module.tex
\section{Tracking Detection}
\label{sec:detection_module}

We detect evasive trackers by looking for a signature that survives identifier rotation and changes in transmission frequency: a \emph{persistently re-occupied} and \emph{abnormally compact} region in CFO-feature space that accumulates evidence across many short-lived identifiers. The physical constraint is that, while logical identifiers can change arbitrarily fast, the analog emitter imprint (captured by CFO statistics) remains sufficiently stable to concentrate the tracker’s segments into a dense region that reappears across windows.

\vspace{0.2em}
\noindent\textbf{Notation.}
Each packet $k$ has timestamp $t_k\!\in\!\mathbb{R}_{\ge 0}$, advertiser address $m_k\!\in\!\mathcal{M}$\, payload $p_k$, and CFO feature vector $\mathbf{f}_k\!\in\!\mathbb{R}^{d_0}$ (5-d CFO features). We use windows of length $W>0$ with index $u_k\triangleq\lfloor t_k/W\rfloor$. Let $\textsf{eco}(\cdot)$ map payloads to a coarse ecosystem class (Apple/Google/Tile/Samsung/Unknown). We define a segmentation identifier $d_k = \textsf{dev}(p_k,m_k)$ as $m_k$ for non-Samsung. For Samsung, we extract a payload-derived private identifier (PRIVID) as $d_k$.

\begin{algorithm}[t!]
\caption{\texttt{AirCatch}'s Tracker Detection}
\label{alg:tracker_module_stream_persist_compact}
\small
\KwIn{Packet stream $(t,m,p,\mathbf{f})$; window $W$; block size $B$;
thresholds $T_{\min},T_{\text{gap}},\delta$.}
\KwOut{Adversary-positive flag(s) $\mathcal{A}_j$ per block $j$.}

\BlankLine
Init $j\!\leftarrow\!0$, block start $t_0$, buffer $\mathcal{P}\!\leftarrow\!\emptyset$;\;
Init persistence state $\Pi=(t^{\textsf{start}}=\bot,\ t^{\textsf{end}}=\bot,\ t^{\textsf{last+}}=\bot)$.\;

\While{stream active}{
Receive packet $(t,m,p,\mathbf{f})$.\;
\If{$t < t_0+(j{+}1)B$}{
$\mathcal{P}\leftarrow \mathcal{P}\cup\{(t,m,p,\mathbf{f})\}$\tcp{accumulate block}
\textbf{continue}
}

$t^{\textsf{end}}_j \leftarrow \max\{t_k : (t_k, m_k, p_k, \mathbf{f}_k)\in\mathcal{P}\}$\;

\textbf{(1) Segmentation:}
$u_k\!\leftarrow\!\lfloor t_k/W\rfloor$\; $d_k\!\leftarrow\!\textsf{dev}(p_k,m_k)$\; $c_k\!\leftarrow\!\textsf{eco}(p_k)$\;
group by $s=(u_k,c_k,d_k)$; compute $(\bar{\mathbf{f}}_s,n_s,t_s^{\min},t_s^{\max})$.\;

\textbf{(2) Type-wise clustering:}
for each $c$: $\mathbf{y}_s\!\leftarrow\!\textsc{zscore}(\bar{\mathbf{f}}_s)$; build $\mathbf{x}_s$;
choose $K$ and Ward-cluster $\{\mathbf{x}_s\}$ into clusters $\tau$ with segments $\mathcal{S}_\tau$.\;

\textbf{(3) Core density:}
\ForEach{cluster $\tau$}{
medoid $s^\star$\; $d_s\!\leftarrow\!\|\mathbf{y}_s-\mathbf{y}_{s^\star}\|_2$\;
$\textsf{Core}(\tau)\!=\!\{d_s\!\le\!\mu_\tau+\lambda\sigma_\tau\}$\;
$\tilde r^{\textsf{core}}_\tau\!\leftarrow\!\max(\mu^{\textsf{core}}_\tau+\lambda\sigma^{\textsf{core}}_\tau,\ r_{\min})$\;
$\textsf{coreDens}_\tau\!\leftarrow\!\frac{(M_\tau/|\mathcal{S}_\tau|)}{\tilde r^{\textsf{core}}_\tau+\epsilon}$.\;
$t^{\min}_\tau \leftarrow \min_{s\in\mathcal{S}_\tau} t^{\min}_s$\;
$t^{\max}_\tau \leftarrow \max_{s\in\mathcal{S}_\tau} t^{\max}_s$\;
}

\textbf{(4) Adversary detection:}
$\mathcal{A}_j\leftarrow\emptyset$\;

$\textsf{pos}_j \leftarrow \exists \tau:\ \textsf{coreDens}_\tau\ge\delta$\;

\If{$\textsf{pos}_j$}{
$t^{\min}_{+}\leftarrow \min_{\tau:\textsf{coreDens}_\tau\ge\delta}\ t^{\min}_\tau$\;
$t^{\max}_{+}\leftarrow \max_{\tau:\textsf{coreDens}_\tau\ge\delta}\ t^{\max}_\tau$\;

\If(\tcp{initialize episode}){$\Pi.t^{\textsf{start}}=\bot$}{
$\Pi.t^{\textsf{start}}\leftarrow t^{\min}_{+}$\;
$\Pi.t^{\textsf{end}}\leftarrow t^{\max}_{+}$\;
}
\Else{
$\Pi.t^{\textsf{start}}\leftarrow \min(\Pi.t^{\textsf{start}},\ t^{\min}_{+})$\;
$\Pi.t^{\textsf{end}}\leftarrow \max(\Pi.t^{\textsf{end}},\ t^{\max}_{+})$\;
}
$\Pi.t^{\textsf{last+}}\leftarrow t^{\max}_{+}$\;

\If{$\Pi.t^{\textsf{end}}-\Pi.t^{\textsf{start}} \ge T_{\min}$}{
$\mathcal{A}_j\leftarrow \{\textsf{ADV}\}$\tcp{adversary-positive}
$\Pi.t^{\textsf{start}}\leftarrow \bot$;\ $\Pi.t^{\textsf{end}}\leftarrow \bot$;\ $\Pi.t^{\textsf{last+}}\leftarrow \bot$
}
}
\Else{
\If{$\Pi.t^{\textsf{last+}}\neq\bot$ \textbf{and} $t^{\textsf{end}}_j-\Pi.t^{\textsf{last+}} > T_{\text{gap}}$}{
$\Pi.t^{\textsf{start}}\leftarrow \bot$;\ $\Pi.t^{\textsf{end}}\leftarrow \bot$;\ $\Pi.t^{\textsf{last+}}\leftarrow \bot$
}
}
$j\leftarrow j{+}1$\; $\mathcal{P}\leftarrow\emptyset$\tcp{advance to next block}
}
\end{algorithm}

\vspace{0.2em}
\noindent\textbf{Step 1: Segmentation (anti-contamination).}
Naive window aggregation in crowded environments pools unrelated emitters and makes CFO averageable, destroying compactness/density geometry and inflating false positives. We therefore segment at
\begin{equation}
s \triangleq (u,\ \textsf{eco}(p),\ d),\qquad
u=\lfloor t/W\rfloor,\quad d=\textsf{dev}(p,m),
\end{equation}
so each segment summarizes \emph{one} device-id within \emph{one} window ($W=120s$) and ecosystem class. This is adversarially robust: (i) transmission duty-cycling cannot hide inside a pooled window; (ii) fast rotation increases the number of segments from the same physical emitter, strengthening persistence and density evidence. For each segment $s$ with packet indices $\mathcal{K}(s)$:
\begin{equation}
\bar{\mathbf{f}}_s \triangleq \frac{1}{|\mathcal{K}(s)|}\sum_{k\in\mathcal{K}(s)} \mathbf{f}_k,\qquad
n_s\triangleq |\mathcal{K}(s)|,\qquad
(t_s^{\min},t_s^{\max}).
\end{equation}

\vspace{0.2em}
\noindent\textbf{Step 2: Type-wise embedding and clustering.}
We cluster segments \emph{separately per ecosystem/type} to avoid cross-type mixing. Within each type, we standardize CFO means:
\begin{equation}
\mathbf{y}_s \triangleq \textsc{zscore}(\bar{\mathbf{f}}_s),
\end{equation}
and form an embedding $\mathbf{x}_s$ that augments $\mathbf{y}_s$ with robust deviation features relative to the type-level median CFO vector $\tilde{\mathbf{c}}$:
\begin{equation}
d^{(2)}_s \triangleq \|\bar{\mathbf{f}}_s-\tilde{\mathbf{c}}\|_2,\qquad
d^{(1)}_s \triangleq \|\bar{\mathbf{f}}_s-\tilde{\mathbf{c}}\|_1.
\end{equation}
We then choose $K$ via silhouette on a small bounded grid and apply Ward agglomerative clustering on $\{\mathbf{x}_s\}$, producing CFO-consistent regions $\tau$, while not striving for perfect device separation. 

\vspace{0.2em}
\noindent\textbf{Step 3: Core selection.} 
For each cluster, we do the core selection and radius computation in the standardized CFO space $\mathbf{y}_s$ (not the full embedding $\mathbf{x}_s$), ensuring the compactness score reflects analog CFO geometry. Let $n_\tau=|\mathcal{S}_\tau|$. We compute the CFO-only medoid
\begin{equation}
s^\star \triangleq \arg\min_{s\in\mathcal{S}_\tau}\ \sum_{s'\in\mathcal{S}_\tau}\|\mathbf{y}_s-\mathbf{y}_{s'}\|_2,
\end{equation}
and distances $d_s\triangleq \|\mathbf{y}_s-\mathbf{y}_{s^\star}\|_2$. The code defines a dense core via a stable dispersion gate:
\begin{equation}
\mu_\tau \triangleq \frac{1}{n_\tau}\sum_{s\in\mathcal{S}_\tau} d_s,\quad
\sigma_\tau \triangleq \sqrt{\frac{1}{n_\tau}\sum_{s\in\mathcal{S}_\tau}(d_s-\mu_\tau)^2},
\end{equation}
\begin{equation}
\textsf{Core}_\lambda(\tau)\triangleq \{\, s\in\mathcal{S}_\tau:\ d_s \le \mu_\tau + \lambda\sigma_\tau \,\},
\end{equation}
with a minimum core size $k_{\min}$ enforced by taking the $k_{\min}$ closest segments to $s^\star$ if needed. The effective core radius is computed within the selected core and clamped away from zero:
\begin{equation}
\tilde r^{\textsf{core}}_\tau \triangleq \max\!\Bigl(\mu^{\textsf{core}}_\tau + \lambda\sigma^{\textsf{core}}_\tau,\ r_{\min}\Bigr),
\end{equation}
where $\mu^{\textsf{core}}_\tau,\sigma^{\textsf{core}}_\tau$ are the mean/std of $\{d_s\}$ restricted to the core. We define identifier diversity and density as
\begin{equation}
\textsf{macDiv}_\tau \triangleq \frac{M_\tau}{n_\tau},\qquad
\textsf{coreDens}_\tau \triangleq \frac{\textsf{macDiv}_\tau}{\tilde r^{\textsf{core}}_\tau+\epsilon}.
\end{equation}
We employ at least two unique IDs to prevent trivially tight single-identifier clusters. We choose $r_{\min}=0.15$ and $\lambda=1.5$. 

\vspace{0.2em}
\noindent\textbf{Step 4: Tracker detection.}
We process packets as a stream in fixed \emph{periodic blocks} of duration $B$ (e.g., $B{=}2400$s) and run Steps~1--3 independently on each block to obtain per-block clusters and their compactness scores. For block $j$, we first form a \emph{block-level} positive indicator:
\begin{equation}
\textsf{pos}_j \triangleq \exists\,\tau \ \text{s.t.}\ \textsf{coreDens}_{\tau,j}\ge \delta .
\label{eq:block_pos_rule}
\end{equation}

We aggregate evidence over a \emph{persistence episode} that tracks the earliest start time and latest end time among all positive clusters observed so far (and discards stale evidence after a gap). Concretely, when $\textsf{pos}_j=\textsf{True}$, we compute the time-span covered by the positive clusters within block $j$:
\begin{equation}
t^{\min}_{+,j} \triangleq \min_{\tau:\ \textsf{coreDens}_{\tau,j}\ge\delta} t^{\min}_{\tau,j},
\qquad
t^{\max}_{+,j} \triangleq \max_{\tau:\ \textsf{coreDens}_{\tau,j}\ge\delta} t^{\max}_{\tau,j},
\end{equation}
where $t^{\min}_{\tau,j}=\min_{s\in\mathcal{S}_{\tau,j}} t^{\min}_{s}$ and $t^{\max}_{\tau,j}=\max_{s\in\mathcal{S}_{\tau,j}} t^{\max}_{s}$ are the cluster time bounds induced by the segment bounds. We then update the episode state by maintaining
$t^{\textsf{start}} \leftarrow \min(t^{\textsf{start}}, t^{\min}_{+,j})$ and
$t^{\textsf{end}} \leftarrow \max(t^{\textsf{end}}, t^{\max}_{+,j})$,
and record the last observed positive time
$t^{\textsf{last+}} \leftarrow t^{\max}_{+,j}$.

At the end of each block, we declare an adversary if the episode’s accumulated time span exceeds $T_{\min}$:
\begin{equation}
t^{\textsf{end}} - t^{\textsf{start}} \ \ge\ T_{\min}.
\label{eq:final_step4_rule_periodic_block}
\end{equation}
To \emph{forget} transient false positives, if a block is negative ($\textsf{pos}_j=\textsf{False}$) we reset the episode whenever the time since the last positive exceeds a gap threshold $T_{\text{gap}}$ ($T_{\text{gap}}$ = 1 day):
\begin{equation}
t^{\textsf{end}}_{j} - t^{\textsf{last+}} \ >\ T_{\text{gap}},
\end{equation}
where $t^{\textsf{end}}_{j}$ is the end time of the just-processed block, i.e., the maximum packet timestamp observed in that block. Finally, once an adversary is flagged, we reset the episode state to avoid repeatedly alerting the same adversary. 

We choose $B{=}2400$s (40~min) as a practical compromise: it is long enough to accumulate sufficient segments per block for stable core-density estimates, while keeping detection latency bounded and amortizing computation in real-time deployment. Users can tune sensitivity via $(T_{\min},T_{\text{gap}})$: smaller values increase responsiveness (including to sparse, duty-cycled adversaries), while larger values reduce spurious alerts by requiring longer sustained evidence. 

%% file: Sections/Designing_an_Advanced_Tracker.tex
\section{Anti-Tracking Device: \blephasyr}
\label{sec:device}

\begin{figure}[t]
    \centering
    \includegraphics[width=\linewidth]{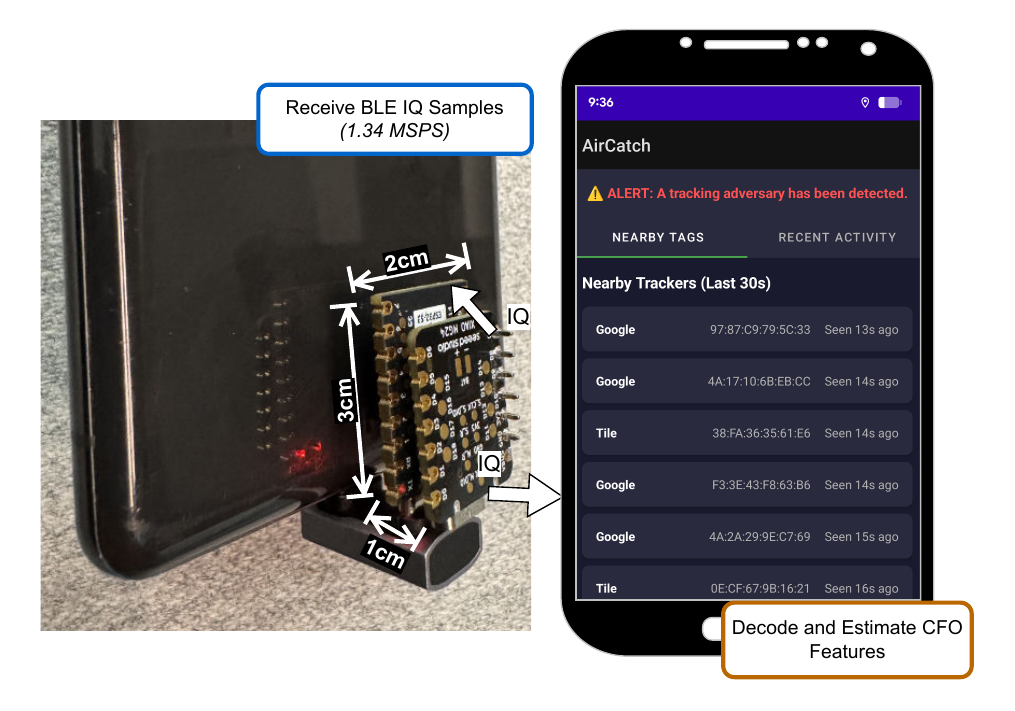}
    \vspace{-0.4em}
    \caption{\textbf{\blephasyr.} The BLE micro-SDR receives IQ data and sends triggered IQ bursts to a host for BLE decoding and CFO estimation. The Android app presents an end-user alert and optional technical views of nearby tag activity.}
    \label{fig:aircatch_device}
    \vspace{-0.8em}
\end{figure}

\texttt{AirCatch} is implemented as a pocket-size, \emph{$\approx$\$10} SDR-based BLE \emph{micro-SDR} named \blephasyr. Our prototype uses a commodity \emph{Seeed XIAO MG24} (Silicon Labs EFR32xG24)\footnote{\url{https://www.seeedstudio.com/Seeed-Studio-XIAO-MG24-p-6247.html}} as the RF front-end and a low-cost \emph{ESP32-S3} companion module\footnote{\url{https://www.alibaba.com/product-detail/ESP32-S3-Development-Board-Super-Mini_1601633352134.html}}. The MG24 exposes raw receive IQ via the \texttt{RAIL} radio interface and streams it to a host over high-speed SPI. Figure~\ref{fig:aircatch_device} shows the complete pipeline and end-to-end UX: \blephasyr \xspace continuously monitors BLE advertising channels and collects \emph{IQ Samples at 1.34\,MSPS}, buffers samples via DMA into a ring, and \emph{SPI-streams} triggered IQ bursts to a host where we run BLE decoding and CFO estimation. In our implementation, the MG24 streams IQ over SPI to a small companion ESP32-S3 (used as a simple, high-throughput transport), which then relays the data to the phone over a wired \mbox{USB-C} link . 

The companion Android application provides a user-facing safety interface: it surfaces a clear \emph{tracking alert} when \texttt{AirCatch} flags an adversary, while also offering an “inspect” view that lists nearby tag activity (e.g., per-ecosystem sightings, recency, and identifiers when visible) to support transparency and debugging without requiring any specialized RF expertise.

On the firmware side, \blephasyr \xspace solves the main physical bottleneck that makes “RF-fingerprinting-grade” sensing difficult on ultra-low-cost devices: sustaining high-throughput IQ capture without overflowing the RX FIFO or stalling the radio. Concretely, the implementation uses (i) a large multi-chunk ring buffer ($\approx$240\,KB of staging) to absorb bursty BLE activity, (ii) tight critical-section queueing to maintain lock-free producer/consumer behavior between the RX FIFO handler and the SPI transport, and (iii) a \emph{polling} LDMA-driven SPI transmit path that minimizes IRQ overhead while sustaining a multi-Mbps stream (13\,MHz EUSART/SPI clock).

To avoid streaming noise-dominated, low-information samples and to maintain practical power and throughput, \blephasyr \xspace employs an energy-triggered capture state machine with a short pre/post history window: it computes a lightweight IQ energy statistic on each chunk and only transmits chunks that exceed a self-tuned threshold, with bounded pre/post capture to retain synchronization context. The firmware further hardens continuous operation with explicit overflow recovery and per-chunk metadata stamping (seed + checksums) to enable end-to-end integrity checks on the host. Together, these engineering choices let a commodity microcontroller reliably deliver the raw physical-layer evidence (\emph{stable CFO features}) required by \texttt{AirCatch} in realistic and dense BLE environments, while keeping the overall system deployable as a low-cost anti-tracking device. 

%% file: Sections/Results.tex
\section{Evaluation}
\label{sec:results}
\begin{figure*}[t]
    \centering
    \begin{subfigure}[t]{0.49\textwidth}
        \centering
        \includegraphics[width=\linewidth]{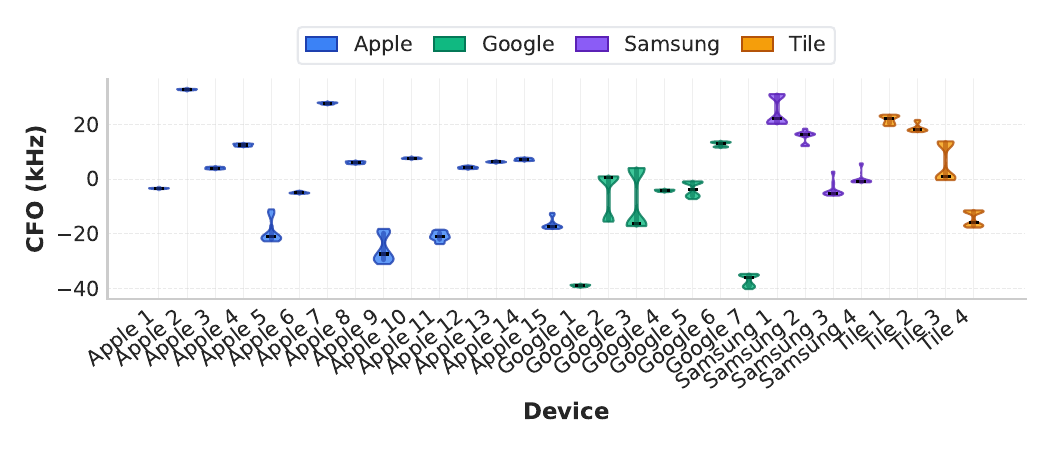}
        \caption{\textbf{USRP B210 (SDR):} per-device CFO distributions for commodity tags. Each device exhibits a tight, device-specific CFO offset with clear inter-device separation.}
        \label{fig:airtag_cfo_b210}
    \end{subfigure}\hfill
    \begin{subfigure}[t]{0.49\textwidth}
        \centering
        \includegraphics[width=\linewidth]{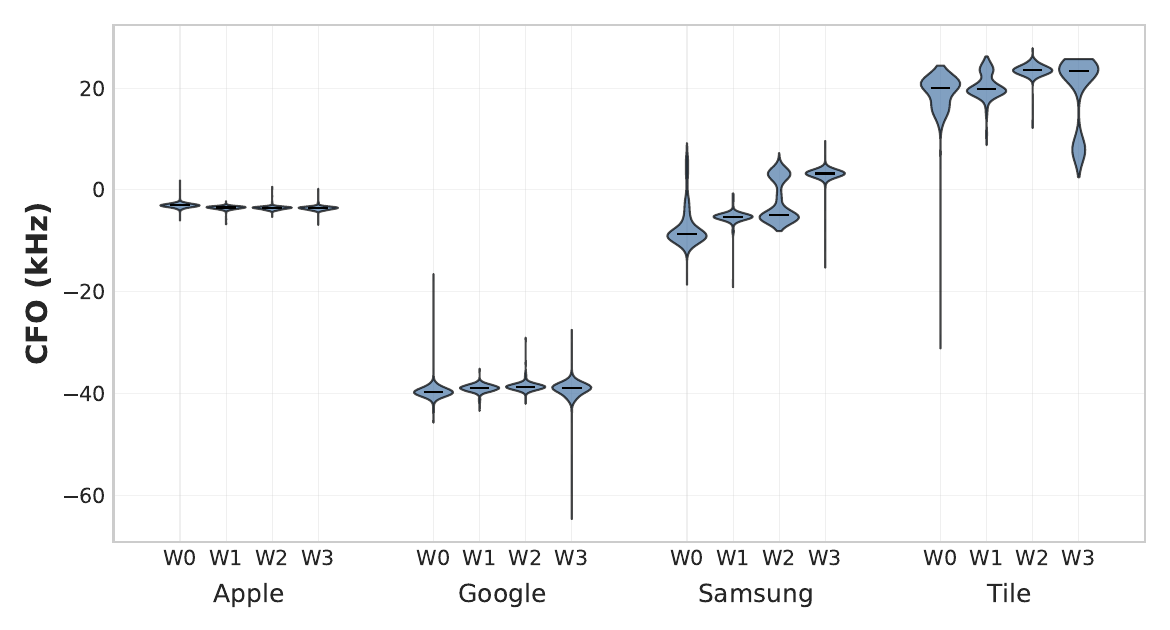}
        \caption{\textbf{USRP B210 (SDR):} CFO over time (four 15-minute windows) for a representative tag of all categories. Consistency across windows indicates limited temporal drift during short periods.}
        \label{fig:airtag_cfo_drift_15min}
    \end{subfigure}
    \vspace{-0.6em}
    \caption{Stability of CFO-based fingerprints under SDR capture.}
    \label{fig:airtag_cfo_stability_b210}
    \vspace{-0.8em}
\end{figure*}
This section presents a comprehensive evaluation of our tracker-detection pipeline across (i) \emph{fingerprint validity} (stability and discriminative power), (ii) \emph{baseline detection} against a naive adversary, and (iii) \emph{evasion robustness} against an advanced adversary that jointly manipulates transmission rate and identifier rotation. Throughout, we emphasize \emph{operational} performance and demonstrate the clustering nature.  

\begin{table}[t]
\centering
\small
\setlength{\tabcolsep}{4.5pt}
\begin{tabular}{lrrr}
\toprule
\textbf{Scenario} & \textbf{Dur.} & \textbf{\#Packets} & \textbf{\#MACs} \\
\midrule
Home$\rightarrow$Work & 113.0\,min & 18{,}216 & 6{,}251 \\
Work$\rightarrow$Home & 104.8\,min & 15{,}513 & 5{,}841 \\
Car commute           & 63.3\,min  & 12{,}315 & 4{,}021 \\
Airport (no adv.)     & 175.5\,min & 8{,}905  & 2{,}467 \\
\midrule
\textbf{Total} & \textbf{456.6\,min} & \textbf{54{,}949} & \textbf{18{,}580} \\
\bottomrule
\end{tabular}
\vspace{1mm}
\caption{Dataset summary. The first three scenarios include an adversary, while the airport scenario is adversary-absent and used to stress-test false positives in dense public settings.}
\label{tab:datasets_summary}
\vspace{-0.8em}
\end{table}

\subsection{Captured Datasets and Environments}
\label{sec:datasets}
\paragraph{Hardware.}
Unless otherwise stated, our adversary devices are custom ESP32-based trackers running modified OpenHaystack firmware. We evaluate two receiver front-ends: (i) \textbf{USRP B210} for high-fidelity IQ capture and CFO extraction, and (ii) \textbf{\blephasyr} directly extracts embedded CFO features on-device. We report results separately per receiver to quantify robustness across sensing stacks.

\tightvspace
\paragraph{Ethics and privacy.}
Our data collection is designed to minimize privacy risk and to avoid collecting personally identifiable information (PII). We capture only over-the-air BLE advertising traffic (public beacons) and do not collect user content, account identifiers, location traces, or any auxiliary metadata beyond what is required to compute our RF features. We anonymize identifiers at ingestion: MAC addresses are \emph{HMACed} and truncated using a fresh, per-scenario secret key, which is discarded after processing, preventing linkage across scenarios or datasets. For Samsung tags, we similarly HMAC-and-truncate the extracted PRIVID solely to obtain per-device ground truth during evaluation; we never retain raw PRIVIDs. Although packet bytes are parsed to locate fields needed for CFO extraction, we do not store raw payloads, and we do not retain any raw IQ samples: both payload and IQ are discarded immediately after feature computation, and only CFO-derived features are retained for analysis. We label as “adversary” only the devices we deploy and control for experiments; all other observed devices are treated as incidental bystanders/background. We conduct captures in public or controlled environments under our supervision and do not attempt to target specific individuals or to associate observed traffic with persons; any bystander traffic is processed only in anonymized, aggregate form. All released artifacts exclude raw identifiers, payloads, and IQ.

\begin{figure*}[t]
    \centering
    \begin{subfigure}[t]{0.49\textwidth}
        \centering
        \includegraphics[width=\linewidth]{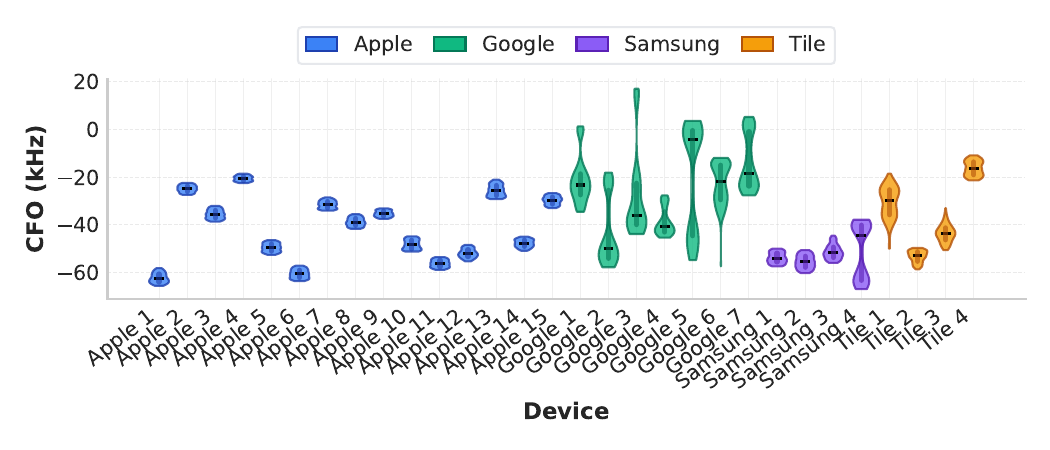}
        \caption{\textbf{\blephasyr:} per-device CFO distributions for commodity tags. CFO offsets and separations closely match the SDR observation in Figure~\ref{fig:airtag_cfo_b210}.}
        \label{fig:all_devices_cfo_rail}
    \end{subfigure}\hfill
    \begin{subfigure}[t]{0.49\textwidth}
        \centering
        \includegraphics[width=0.55\linewidth]{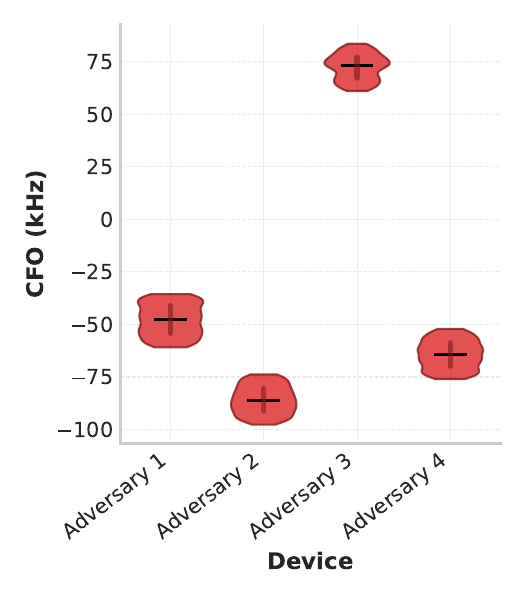}
        \caption{\textbf{\blephasyr:} CFO distributions for four ESP32-based advanced adversary trackers. Devices exhibit distinct CFO medians, supporting stable accumulation under per-transmission pseudonym rotation.}
        \label{fig:adv_devices_cfo_rail}
    \end{subfigure}
    \vspace{-0.6em}
    \caption{\blephasyr \xspace validation: CFO stability/separation on commodity tags, and separability of advanced adversary devices.}
    \label{fig:cfo_micro_sdr}
    \vspace{-0.8em}
\end{figure*}

\tightvspace
\paragraph{Datasets.}
We evaluate \texttt{AirCatch} on four real-world mobility traces captured with our low-cost \emph{micro-SDR} (\blephasyr) on channel 37. Each trace contains multiple mobility contexts (e.g., walking, transit segments, indoor stops); we use the context boundaries recorded in the input CSV to support phase-wise analysis and background shading in time-series plots. The first three traces include an \emph{adversary-present} condition (our ESP32-based advanced tracker co-located with and following the user), whereas the airport trace is collected \emph{without} an adversary as a stress test for false positives in a dense public setting (notably the baggage reclaim area). Table~\ref{tab:datasets_summary} summarizes each dataset; in total, our mobility evaluation spans \textbf{456.6 minutes}, comprising \textbf{54{,}949 packets} and \textbf{18{,}580 observed MAC addresses}. Moreover, in this paper, we consider four representative tracker ecosystems: Apple AirTag, Google Find My tags, Tile, and Samsung SmartTag for the analysis of commodity tags.  

\tightvspace
\paragraph{Adversarial scenario generation.}
We evaluate an advanced adversary who explicitly attempts to evade detection by tuning its effective advertising period $T_{\mathrm{tx}}$. In our threat model, the adversary \emph{rotates its identifier at every transmission}; hence the identifier rotation period is \emph{coupled} to the transmission period, i.e., $T_{\mathrm{rot}} = T_{\mathrm{tx}}$. This attacker is realistic and requires only firmware control over advertising rate and address randomization. We generate adversarial datasets using our scenario generator by sweeping
\[
T_{\mathrm{tx}} \in \{2,10,15,30,60\}\,\text{s},
\qquad \text{with } T_{\mathrm{rot}} = T_{\mathrm{tx}}.
\]
In the baseline traces of Table~\ref{tab:datasets_summary}, both the adversary and benign tags transmit at a nominal advertising interval of 2s. For each $T_{\mathrm{tx}}$ setting, we synthesize an evading adversary by replaying the selected tag’s packets after (i) downsampling to the target transmission period and (ii) rotating its pseudonym on every emission. Varying $T_{\mathrm{tx}}$ thus spans adversaries from highly active to stealthy.

\tightvspace
\paragraph{Stealth--utility trade-off.}
This coupling induces a fundamental trade-off for the adversary. At \emph{high transmission frequency} (small $T_{\mathrm{tx}}$), the attacker emits more packets, which increases detection risk by producing denser and more persistent CFO-consistent evidence. However, frequent transmissions also improve tracking utility: more advertisements yield more opportunities for nearby third-party devices to observe the tracker and upload location reports, increasing the temporal granularity and reliability of victim localization. Conversely, in \emph{stealth mode} (large $T_{\mathrm{tx}}$), the attacker reduces its RF footprint and weakens the evidence available to the detector, but necessarily sacrifices tracking performance by lowering the rate of location updates and increasing gaps between observations.

We implemented an adversarial AirTag using a modified version of OpenHaystack \cite{heinrich2021openhaystack}. The device transmits a single packet and then rotates its MAC address every two seconds. We subsequently queried Apple’s Find My network for corresponding reports and observed successful detections for most ephemeral keys. This indicates that a single transmission can suffice for tracking, although the success probability is lower.




\subsection{Effectiveness of Chosen Fingerprints}
\label{sec:results:fingerprints}

We first validate that our CFO-derived fingerprints satisfy two necessary properties for robust tracker detection under identifier rotation:
\textbf{(i) stability} within a device over time and across contexts, and
\textbf{(ii) discriminative} nature across devices (so that rotated pseudonyms still accumulate into a compact, device-consistent CFO core).
We evaluate both properties on (a) a high-fidelity USRP B210 SDR pipeline and (b) our low-cost ESP32+RAIL, which extracts the same CFO and transition-CFO features on-device.

\subsubsection{Stability across time and sensing stacks}
\label{sec:results:stability}

\paragraph{Protocol.}
We record $\approx$1-hour traces with multiple commodity tags per ecosystem operating under nominal behavior. We compute per-packet CFO and aggregate per-device distributions. To quantify time stability, we partition each trace into 15-minute windows and compare each device’s CFO distribution across windows (the same windowing granularity later used by our detection module).

\tightvspace
\paragraph{CFO stability (B210 SDR).}
Figure~\ref{fig:airtag_cfo_stability_b210} (left) shows per-device CFO distributions for commodity tags captured using the USRP B210. Each device exhibits a \emph{tight, device-specific} CFO offset, while different devices occupy distinct CFO regions. Across all tags shown, CFO offsets span only \emph{tens of kHz} (approximately within $[-80, +40]$\,kHz in this capture), and the per-device spreads are narrow relative to inter-device separation, exactly the regime needed for CFO-based clustering to remain coherent even if identifiers rotate.

\tightvspace
\paragraph{Limited temporal drift (15-minute windows).}
Figure~\ref{fig:airtag_cfo_stability_b210} (right) reports the same devices partitioned into four consecutive 15-minute windows. Device CFO distributions remain highly consistent across windows: the medians remain near-constant and the within-window spreads overlap substantially. In practice, this bounded drift supports (i) windowed aggregation in our pipeline, and (ii) long-horizon persistence tests (Section~\ref{sec:results:multiscenario}), because a tracker must remain inside a compact CFO core for long periods to be flagged.

\tightvspace
\paragraph{\blephasyr \xspace agreement with SDR and adversary CFO separation.}
Figure~\ref{fig:cfo_micro_sdr} validates that \blephasyr \xspace produces CFO fingerprints that are \emph{qualitatively consistent} with the SDR pipeline: device CFO offsets again lie within a narrow tens-of-kHz band and preserve inter-device separation (left). This is critical because our \emph{deployment} setting relies on the \blephasyr. Further, Figure~\ref{fig:cfo_micro_sdr} (right) shows CFO distributions for our ESP32-based \emph{advanced adversary} trackers as observed by the \blephasyr. The adversary devices occupy distinct CFO regions (with medians separated by \emph{multiple} kHz-to-tens-of-kHz), enabling their rotated pseudonyms to consistently accumulate into a compact CFO core rather than dispersing into background clusters.

\begin{figure}[t]
    \centering
    \includegraphics[width=\linewidth]{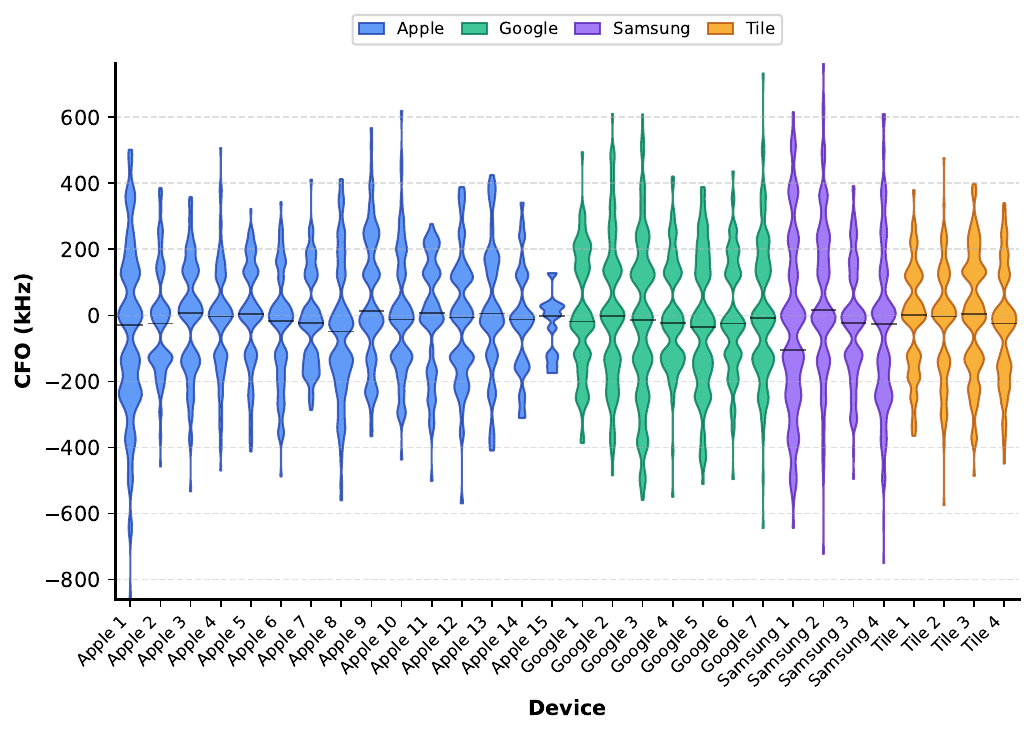}
    \vspace{-0.6em}
    \caption{CFO estimates from prior CFO-based fingerprinting ~\cite{cfo_ucsd}. Outputs show high variance, contrasting with the stable tens-of-kHz regime of our CFO/transition-CFO pipeline.}
    \label{fig:ucsd_cfo}
    \vspace{-0.8em}
\end{figure}

\subsubsection{Discriminative Nature of CFO Features}
\label{sec:results:discriminative}

\begin{figure}[t]
    \centering
    \includegraphics[width=1.0\linewidth]{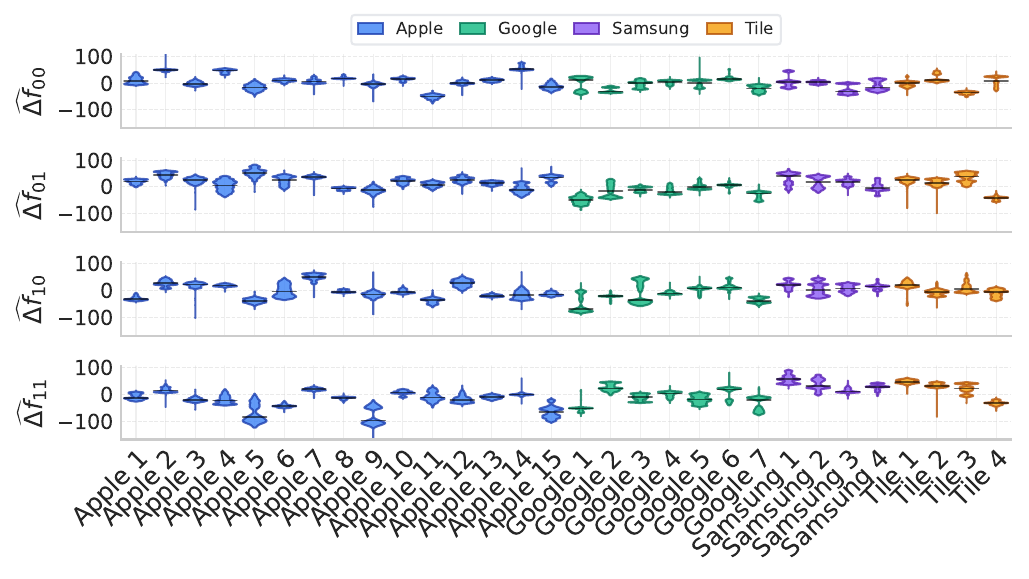}
    \vspace{-0.6em}
    \caption{Transition-CFO fingerprints ($\Delta f_{00}$, $\Delta f_{01}$, $\Delta f_{10}$, $\Delta f_{11}$) in kHz for commodity tags. Transition-conditioned CFO features remain stable per device and provide additional separation beyond a single CFO offset.}
    \label{fig:transition_cfo}
    \vspace{-0.68em}
\end{figure}

\paragraph{Transition-CFO features add structure beyond a single offset.}
A single CFO offset can already separate many devices, but in crowded environments (or under partial overlap of CFO medians) we further leverage \emph{transition-CFO} features, which capture CFO conditioned on the instantaneous symbol transitions (e.g., $\widehat{\Delta f}_{00}$, $\widehat{\Delta f}_{01}$, $\widehat{\Delta f}_{10}$, $\widehat{\Delta f}_{11}$). Figure~\ref{fig:transition_cfo} shows that these transition-conditioned distributions remain \emph{device-consistent} while adding complementary axes of separation, increasing discriminability when plain CFO alone is ambiguous. This directly benefits our clustering stage: rotated identifiers from the same physical tracker remain close in the joint CFO/transition-CFO embedding, while background devices remain dispersed.

\tightvspace
\paragraph{Supervised fingerprinting confirms separability (SDR and \blephasyr).}
To quantify discriminative power, we perform within-ecosystem device identification using only CFO-derived fingerprints. We train a Random Forest classifier to predict device identity from CFO + transition-CFO features using an 80:20 \emph{time} split on a 1-hour capture (train and test are disjoint time blocks to avoid temporal leakage). Results are strong and, critically, consistent across sensing stacks.

\vspace{-0.35em}
\begin{table}[t]
\centering
\footnotesize
\setlength{\tabcolsep}{3.5pt}
\renewcommand{\arraystretch}{1.05}
\begin{tabular}{lccccc}
\toprule
\textbf{Front-end} & \textbf{Acc.} & \textbf{Prec.} & \textbf{Rec.} & \textbf{F1} \\
\midrule
B210 (SDR) & 85.53\% & 84.53\% & 85.53\% & 83.21\% \\
RAIL (\blephasyr) & 85.59\% & 84.30\% & 85.59\% & 83.77\% \\
\bottomrule
\end{tabular}
\vspace{0.5mm}
\caption{Device identification within type using CFO + transition-CFO features (RF; 80:20 time split on a 1-hour trace). The \blephasyr \xspace matches SDR-level discriminability.}
\label{tab:fingerprint_rf}
\end{table}

These results validate that CFO fingerprints are (i) sufficiently separable to link rotated pseudonyms back to the same physical device in feature space, and (ii) robust to a low-cost sensing pipeline, supporting our goal of \emph{deployable} anti-tracking.

\tightvspace
\paragraph{Comparison to prior CFO-based fingerprinting~\cite{cfo_ucsd}.}
 We additionally compared against a prior CFO-based fingerprinting
 module. Due to runtime constraints, we ran it on only 5{,}000 packets, and it produced CFO estimates for $\approx$3{,}400 packets (failures are expected under low SNR / synchronization sensitivity). Figure~\ref{fig:ucsd_cfo} shows that its CFO outputs exhibit markedly higher variance (spanning \emph{hundreds of kHz}, roughly $[-800,+600]$\,kHz in our run) and are substantially less stable than our CFO/transition-CFO pipeline (Figures~\ref{fig:airtag_cfo_stability_b210}--\ref{fig:transition_cfo}). This reinforces a core design choice in \texttt{AirCatch}: we favor CFO fingerprints that are \emph{stable, lightweight, and robust enough to run continuously} on a \blephasyr, rather than high-variance estimators that are difficult to operationalize at scale.

\subsection{Adversary Detection by State-of-the-Art}
\label{sec:results:status_quo}

\paragraph{Detection infrastructure.}
We evaluate tracker detectability using both platform-native protections and third-party tooling. Concretely, we carry an Apple iPad Pro (M1) running iPadOS 16.2 (iOS unwanted-tracker alerts) and a Google Pixel 7 Pro running Android 16 (Android unknown-tracker alerts), and we additionally run \emph{AirGuard} on the both of these devices to provide cross-ecosystem coverage. This setup lets us compare (i) Apple’s and Google’s built-in mechanisms against (ii) a widely used third-party detector. We note that iOS restricts third-party access to BLE MAC addresses, limiting observability for non-system applications and thereby constraining third-party detectors on iOS.

\tightvspace
\paragraph{Threat model and expectation.}
We consider an adversary that deploys a \emph{single physical tracker} while frequently rotating protocol-visible identifiers (e.g., MAC address, public key material, or both depending on ecosystem constraints). Existing deployed defenses largely depend on correlating identifiers and advertisement metadata over time; thus, sufficiently fast rotation and/or sparse transmissions should reduce linkability and delay (or prevent) alerts. Our goal here is not to exhaustively benchmark OS implementations, but to characterize whether current detectors remain effective under realistic evasion patterns used by advanced trackers.

\tightvspace
\paragraph{Naive trackers already incur non-trivial alert latencies.}
In repeated trials where a detector supports the tracker ecosystem, we observe alert latencies of at least 30 minutes and sometimes exceeding one hour. Detection generally occurs faster during nighttime conditions. Among the evaluated systems, AirGuard consistently generated alerts sooner than both Apple and Google while maintaining broader ecosystem coverage. Apple devices detected AirTag variants and also detected Google-based trackers. In contrast, Google devices detected original AirTags and Google tags but did not detect other Find My–compatible trackers from third-party manufacturers. Trackers from Samsung and Tile were not detected by either Apple or Google in our tests. AirGuard, however, detected Apple, Google, Samsung, and Tile trackers, typically within a similar ~30-minute window. 


\tightvspace
\paragraph{Advanced rotation and stealth undermine state-of-the-art detection.}
We then stress detectors with more capable adversaries that rotate identifiers aggressively and/or reduce transmission rate. These settings directly target the correlation assumptions in deployed defenses. We observe that (i) fast-rotating configurations can evade or substantially delay detection across current mechanisms, and (ii) stealthy configurations further degrade detection by reducing packet support, making the tracker appear indistinguishable from transient background devices. In our experiments, cases that combine identifier rotation (within protocol constraints) with reduced transmission opportunities were particularly problematic: detectors either did not surface an alert within the co-location window or produced alerts only after long delays.

\tightvspace
\paragraph{Protocol constraints do not eliminate the threat.}
Apple’s Find My tightly couples public key material to the BLE address in advertisement processing, which restricts certain \emph{MAC-only} or \emph{key-only} rotation modes. However, this coupling does not guarantee timely detection in practice: under combined identifier changes permitted by the ecosystem, built-in mechanisms can still fail to flag a co-moving tracker. Conversely, ecosystems where MAC and identifier rotation are more decoupled inherently expose a broader evasion surface, enabling rapid rotation strategies that directly frustrate identifier-based correlation.

\begin{table*}[t]
\centering
\scriptsize
\setlength{\tabcolsep}{5.0pt}
\renewcommand{\arraystretch}{0.97}
\begin{tabular}{lccccc}
\toprule
\textbf{Scenario} &
\textbf{Benign trace (adv-absent)} &
\textbf{ESP32 adv. (1--4)} &
\textbf{Apple replay adv. (1)} &
\textbf{Google replay adv. (1)} &
\textbf{$T_{\mathrm{tx}}{=}T_{\mathrm{rot}}$ settings} \\
\midrule
Airport (stress test; no adv.) &
\cmark &
-- &
-- &
-- &
-- \\
Home$\rightarrow$Work (adv present) &
\cmark &
\cmark &
\cmark &
\cmark &
$\{2,10,15,30,60\}$\,s \\
Work$\rightarrow$Home (adv present) &
\cmark &
\cmark &
\cmark &
\cmark &
$\{2,10,15,30,60\}$\,s \\
Car commute (adv present) &
\cmark &
\cmark &
\cmark &
\cmark &
$\{2,10,15,30,60\}$\,s \\
\bottomrule
\end{tabular}
\vspace{2pt}
\caption{\texttt{AirCatch}'s tracker detection. Each row is a scenario; the first column reports the \emph{captured benign scenario} (i.e., no adversary present). The remaining columns report whether \texttt{AirCatch} correctly flags adversary-present blocks when we inject (i) 1--4 distinct \emph{ESP32} trackers (each rotating its pseudonym \emph{at every transmission}), and (ii) ``replay'' adversaries created from \emph{commodity Apple/Google tags} by downsampling to a target transmission period and enforcing per-transmission pseudonym rotation. \cmark\ denotes correct behavior (no false alarms on benign traces; correct flag when an adversary is present).}
\label{tab:multiscenario_compact}
\vspace{-0.7em}
\end{table*}

\subsection{Adversary detection by \texttt{AirCatch}}
\label{sec:results:multiscenario}

\begin{figure}[t]
  \centering
  \includegraphics[width=\linewidth]{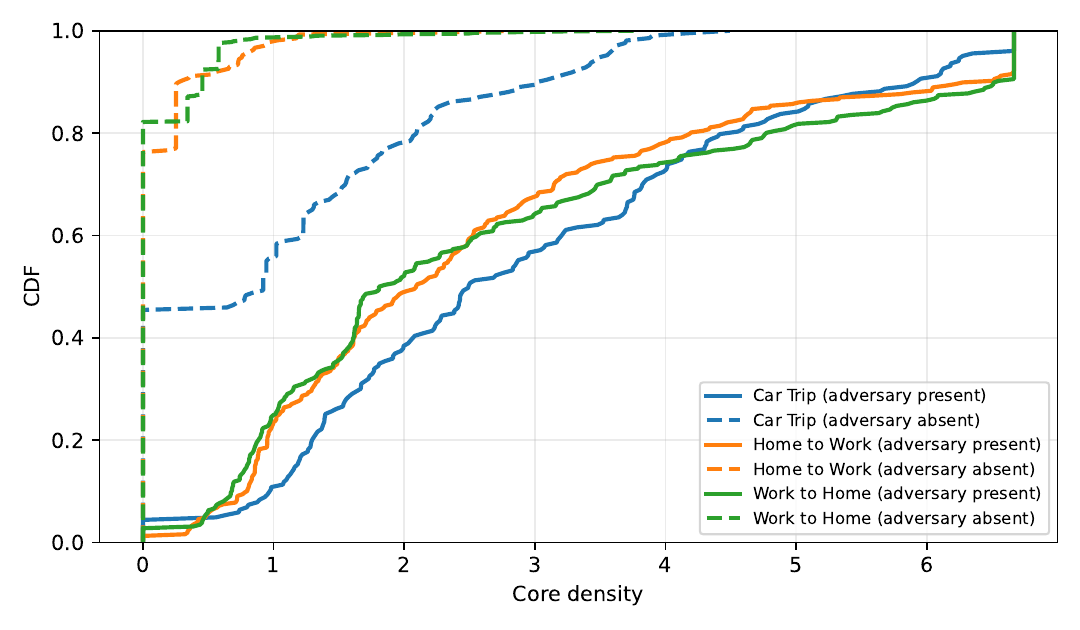}
  \caption{Core-density distribution shift under tracking. Empirical CDF of cluster core densities aggregated per route, overlaying adversary-present vs adversary-absent traces.}
  \label{fig:core_density_cdf}
  \vspace{-0.6em}
\end{figure}

We set the persistence requirement to $T_{\min}=40$ minutes as a practical trade-off between early detection and false-positive robustness. 
This value is chosen to provide timely alerts to users while ensuring that benign co-location events do not accumulate sufficient evidence to trigger detection. 
Lower thresholds enable faster responses but increase the risk of transient dense clusters producing false positives, whereas higher thresholds further suppress false alarms at the cost of delayed adversary detection. We find $40$ minutes to strike a favorable balance between responsiveness and reliability. 

Across multiple scenarios (three real mobility commutes with an adversary present, plus one adversary-absent airport trace), \texttt{AirCatch} cleanly separates background lost-state tag traffic from advanced tracking behavior. 

Concretely, for \emph{all} benign captures (Home$\rightarrow$Work, Work$\rightarrow$Home, Car commute, and Airport stress-test), the detector remains negative throughout, despite high tag-traffic and rapid context transitions (walking, transit, indoor stops). 

We deploy an advanced attacker that \emph{rotates its pseudonym at every transmission} and varies its transmission period $T_{\mathrm{tx}}$, spanning from highly active ($T_{\mathrm{tx}}{=}2$s) to stealthy ($T_{\mathrm{tx}}{=}60$s) operation. We evaluate (i) 1--4 concurrent ESP32-based advanced trackers (distinct physical devices), and (ii) replayed Apple and Google tags made to emulate the same \emph{per-transmission} rotation and downsampled transmission schedule using our scenario generator. Table~\ref{tab:multiscenario_compact} summarizes \texttt{AirCatch}'s detection outcomes, where $\checkmark$ denotes correct behavior (no false alarms on benign traces; correct flags when an adversary is present). Whenever an adversary is present, \texttt{AirCatch} consistently raises an alert: the adversary induces a \emph{persistently compact} CFO core whose density exceeds the operational threshold $\delta$ and whose persistence exceeds $T_{\min}$, while background clusters remain short-lived and typically are below $\delta$.


\begin{figure*}[t]
    \centering
    \begin{subfigure}[t]{0.32\textwidth}
        \centering
        \includegraphics[width=\linewidth]{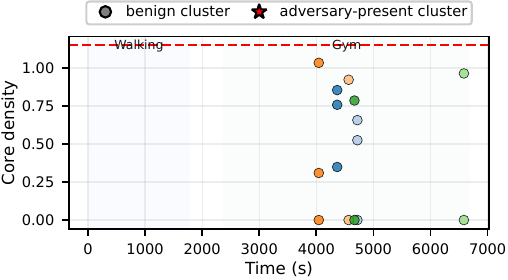}
        \caption{Home$\rightarrow$Work (background only)}
        \label{fig:dens_bg_hw}
    \end{subfigure}\hfill
    \begin{subfigure}[t]{0.32\textwidth}
        \centering
        \includegraphics[width=\linewidth]{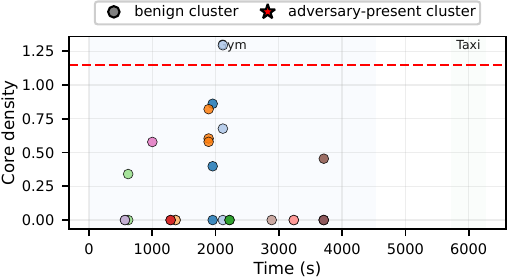}
        \caption{Work$\rightarrow$Home (background only)}
        \label{fig:dens_bg_wh}
    \end{subfigure}\hfill
    \begin{subfigure}[t]{0.32\textwidth}
        \centering
        \includegraphics[width=\linewidth]{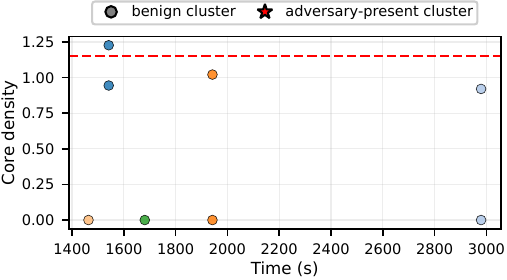}
        \caption{Car commute (background only)}
        \label{fig:dens_bg_car}
    \end{subfigure}

    \vspace{2mm}

    \begin{subfigure}[t]{0.32\textwidth}
        \centering
        \includegraphics[width=\linewidth]{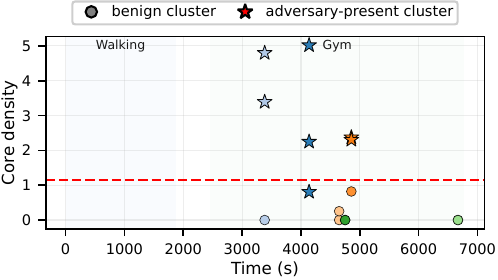}
        \caption{Home$\rightarrow$Work (+ adversary)}
        \label{fig:dens_adv_hw}
    \end{subfigure}\hfill
    \begin{subfigure}[t]{0.32\textwidth}
        \centering
        \includegraphics[width=\linewidth]{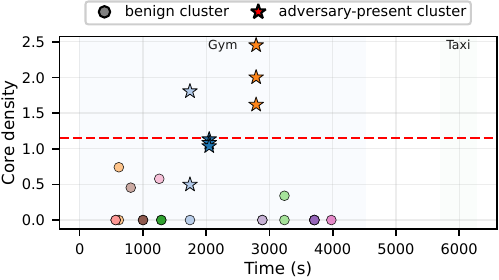}
        \caption{Work$\rightarrow$Home (+ adversary)}
        \label{fig:dens_adv_wh}
    \end{subfigure}\hfill
    \begin{subfigure}[t]{0.32\textwidth}
        \centering
        \includegraphics[width=\linewidth]{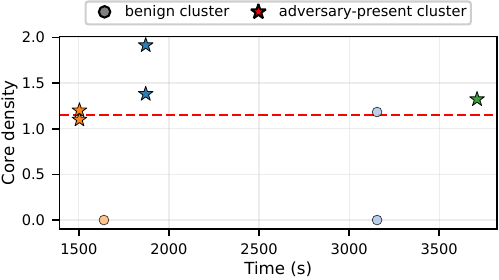}
        \caption{Car commute (+ adversary)}
        \label{fig:dens_adv_car}
    \end{subfigure}

    \caption{Core density over time for three mobility scenarios, shown for (top) background-only traces and (bottom) traces with an ESP32-based advanced adversary (stealth, transmitting per minute with new identifier) following the user. Adversary is flagged only if clusters' density exceeds $\delta=1.15$ (shown by the dashed red line) and they persist for at least $T_{\min}=40$ minutes.}
    \label{fig:dens_vs_time_walkthrough}
    \vspace{-0.8em}
\end{figure*}

\subsection{Behavior of Core Densities}
\label{sec:results:coredensity}

\paragraph{Sensitivity to the core-density threshold $\delta$.}
We stress-test the decision threshold by sweeping $\delta$ around our chosen operating point ($\delta{=}1.15$) and observing the transition from \emph{over-sensitive} (false positives) to \emph{over-strict} (false negatives) regimes. When $\delta$ is reduced below the selected value, the detector begins to admit benign compact clusters: at $\delta{=}0.9$ we observe positives in benign settings (e.g., Airport and Car commute), while true adversaries remain detectable. At $\delta{=}1.0$, the Airport false positive persists, confirming that too-low thresholds primarily hurt precision without improving recall. In contrast, increasing $\delta$ makes the rule conservative and starts to miss adversary blocks: at $\delta{=}1.3$ we already incur false negatives in Car commute, and at $\delta{=}1.45$ the degradation is stronger (e.g., in Work to home and in Car commute). These results validate $\delta{=}1.15$ as an \emph{optimal knee point}: it is the smallest value that suppresses the observed false positives (restoring precision) while still retaining full recall (no false negatives) across our evaluated adversary settings. Practically, $\delta$ provides a user-tunable robustness knob: selecting $\delta{>}\!1.15$ can further harden against rare false positives, but necessarily increases the risk of missed detections for sparse trackers.  

\paragraph{Core-density distribution.}
Figure~\ref{fig:core_density_cdf} shows a pronounced distributional separation between background-only and adversary-present traces.
At our operational threshold $\delta=1.15$, the \emph{background-only} CDF already saturates for two routes: 98.43\% (Work$\rightarrow$Home) and 98.78\% (Car) of cluster cores fall at or below $\delta$, i.e., only 1.57\% and 1.22\% exceed the threshold, respectively.
Home$\rightarrow$Work is the most challenging background route (denser ambient RF activity), yet still 59.18\% of cores remain $\le\delta$ and its high-density tail is handled by the persistence constraint (near-threshold clusters are short-lived and fail $T_{\min}$).
In sharp contrast, in \emph{adversary-present} traces only 14.15\% (Home$\rightarrow$Work), 27.18\% (Work$\rightarrow$Home), and 29.89\% (Car) of cores lie below $\delta$, meaning that 70--86\% of cores exceed the density threshold when tracking is present.
This shift is not a small ``margin'' effect: the median core density moves from $\le 0.92$ in background-only traces to $\approx 1.81$--$2.49$ under tracking, and the upper tail expands dramatically (e.g., 95th-percentile core density reaches 6.27--6.67 with an adversary, versus 0.58--0.76 in two of the three background routes).
Overall, the CDF confirms that core density is a robust route-agnostic discriminator: benign co-location concentrates at low densities, whereas tracking reliably induces a heavy high-density tail, enabling a low-FP operating point when combined with persistence ($T_{\min}$).

\tightvspace
\paragraph{Why core density separates benign co-location from tracking.}
Figure~\ref{fig:dens_vs_time_walkthrough} illustrates the temporal evolution of core density for clusters produced by our pipeline across three real mobility traces (Home$\rightarrow$Work, Work$\rightarrow$Home, and car commute), each segmented into contextual phases (e.g., walking, gym, taxi). In the \emph{background-only} traces (top row), cluster densities remain consistently low and short-lived: Home$\rightarrow$Work never exceeds a density of $1.0$, while the remaining commutes fluctuate within a narrow band below $1.25$, with the vast majority of clusters falling beneath our operational threshold $\delta=1.15$. 
This behavior reflects benign RF environments in which clusters correspond to distinct devices with heterogeneous CFOs and limited temporal overlap, yielding sparse and transient structure rather than a repeatedly accumulating CFO core. 
Potential inflation from continuously transmitting stationary devices is mitigated by our windowed aggregation strategy (Algorithm~\ref{alg:tracker_module_stream_persist_compact}), which prevents packet-level repetition from artificially increasing cluster density.

We further quantify this effect by examining the persistence of dense CFO clusters (core density $>\delta$) under benign and adversarial conditions. 
In benign scenarios, only a small number of clusters ever exceed the density threshold, with $n{=}10$ in the airport stress trace, $n{=}86$ during the car commute, $n{=}5$ for Home$\rightarrow$Work, and $n{=}9$ for Work$\rightarrow$Home. 
In contrast, adversarial scenarios yield substantially more dense clusters, with $n{=}175$, $n{=}224$, and $n{=}201$ clusters in the car commute, Home$\rightarrow$Work, and Work$\rightarrow$Home traces, respectively. 
Moreover, adversarial clusters persist for significantly longer durations, exhibiting median persistence between $1570$\,s and $2120$\,s, often spanning most or all of a detection block, whereas benign clusters remain short-lived, with medians well below this range (e.g., $760$\,s and $1180$\,s). 
Consequently, benign activity rarely accumulates sufficient sustained evidence across blocks to trigger detection, while adversarial trackers consistently form long-lived dense cores that surpass the persistence threshold. 

Consistent with this quantitative behavior, when an ESP32-based evasive adversary is present (bottom row), transmitting intermittently and rotating ADV addresses, the adversary-associated cluster rapidly forms an abnormally compact CFO core whose density rises well above $\delta$ and remains elevated over time. 
In Home$\rightarrow$Work, density reaches values up to $5$, while Work$\rightarrow$Home and the car commute reach approximately $2.5$ and $2.0$, respectively, reflecting repeated accumulation of short-lived identifiers within the same CFO-consistent region. 
While a small number of benign clusters may briefly approach $\delta$ during high-density mobility phases, these fail to satisfy the persistence constraint and decay before $T_{\min}=40$ minutes. 
Together, the conjunction of density and persistence enables near-zero operational false positives while preserving sensitivity to trackers that actively evade linkability by rotating and intermittently transmitting. 


%% file: Sections/Discussion.tex

\section{Limitations}
\label{sec:limitations_cfo_obfuscation}

\texttt{AirCatch} exploits the fact that packets from a single physical tracker concentrate into a persistently dense CFO core despite identifier rotation; this evidence can be weakened if a sophisticated adversary can deliberately \emph{perturb its transmitted carrier frequency} across identifiers or even per transmission (e.g., by tuning the radio synthesizer), thereby spreading its apparent CFO over a wider radius and fragmenting what would otherwise be a compact cluster. In practice, however, such obfuscation is not a free win: it typically requires low-level radio control that may be unavailable or costly on many commodity tags, must remain within receiver tolerances to avoid increasing packet loss (hurting the tracker’s own crowd-reporting effectiveness), and can introduce calibration/temperature-drift management challenges that complicate reliable operation over long deployments. We therefore view CFO obfuscation as a realistic \emph{upper-bound} attacker capability and leave as future work extending \texttt{AirCatch} with CFO-invariant physical features (e.g., transient dynamics or hardware-impairment signatures) so that dense-core evidence can still be accumulated even when CFO is intentionally randomized.

%% file: Sections/appendix.tex
\appendix

\input{Sections/Behavior_of_current_tags} 

\section{CFO features from Ubertooth}
\label{app:ubertooth}

\paragraph{Hamming-weight CFO features from Ubertooth.}

\begin{figure}[t]
    \centering
    \includegraphics[width=\linewidth]{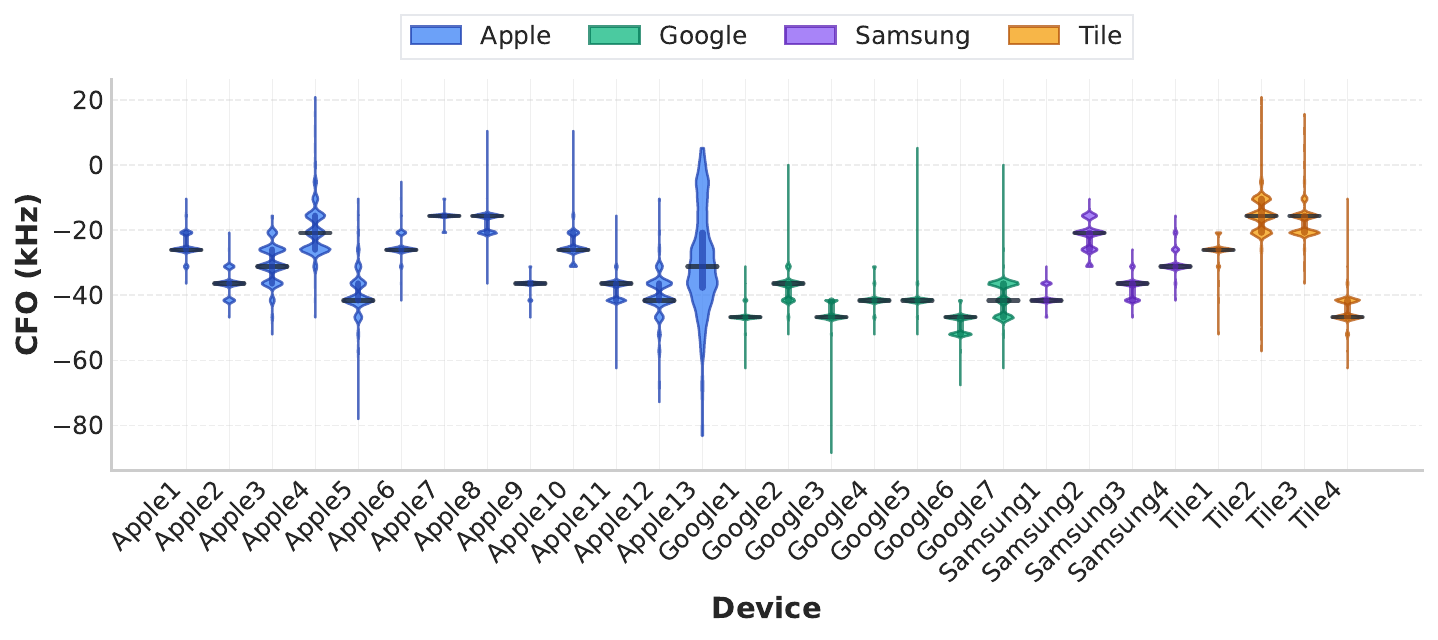}
    \vspace{-0.4em}
    \caption{Ubertooth CFO proxy stability. Per-device distributions of the packet-level \texttt{FREQEST}-derived CFO proxy, showing tight within-device concentration and clear inter-device separation.}
    \label{fig:Uber_cfoTot_violin_persistent}
    \vspace{-0.8em}
\end{figure}

\begin{figure}[t]
    \centering
    \includegraphics[width=\linewidth]{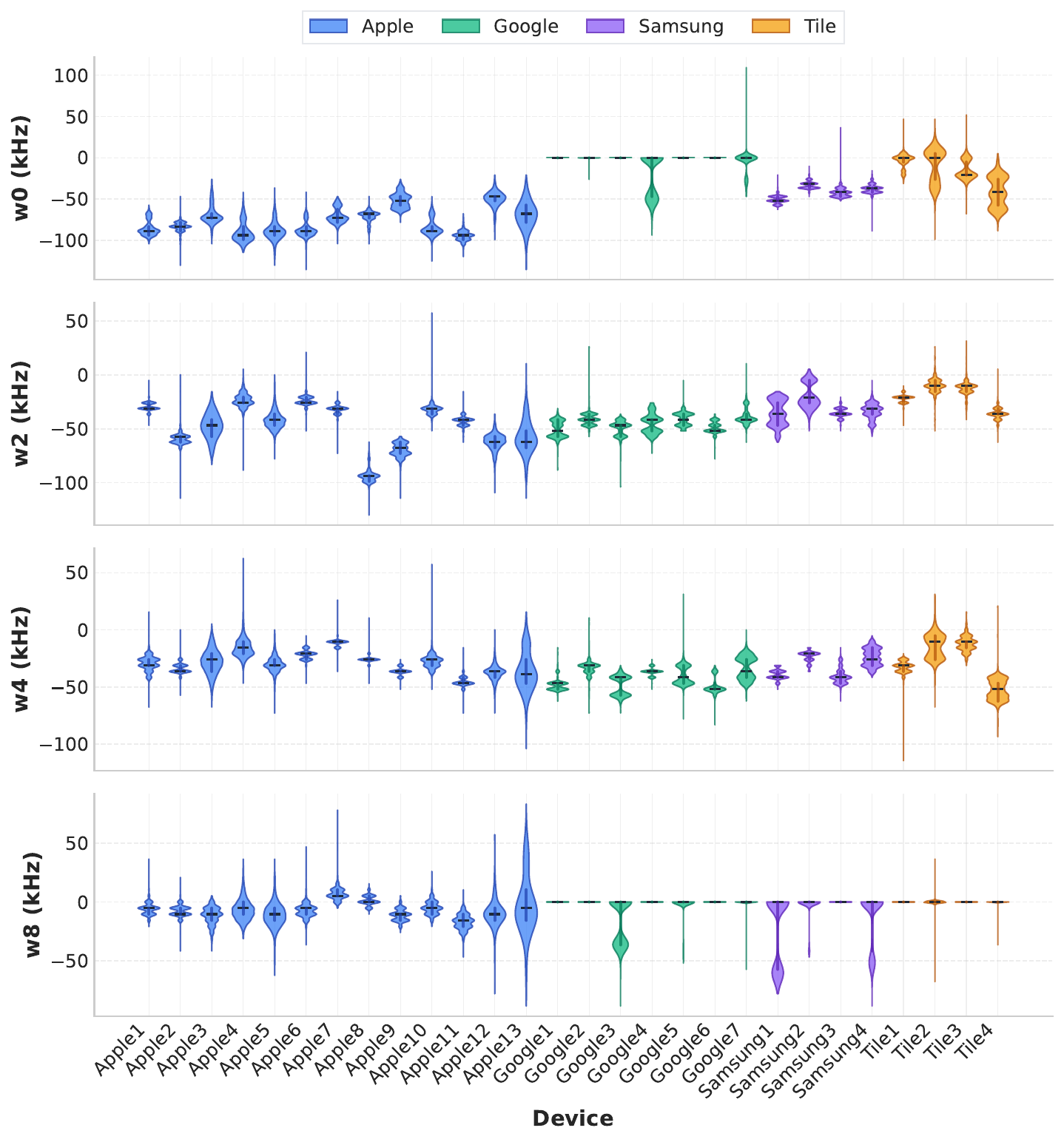}
    \vspace{-0.4em}
    \caption{Hamming-weight–conditioned CFO features (Ubertooth). Conditioned CFO means across representative Hamming-weight bins (e.g., $w\!\in\!\{0,2,4,8\}$) provide additional device-specific evidence beyond a single global CFO proxy.}
    \label{fig:Uber_weighted_features}
    \vspace{-0.8em}
\end{figure}

When capturing with Ubertooth-class devices, we do not have access to full-rate complex IQ samples; instead, the receiver exposes a per-byte frequency estimate, \texttt{FREQEST}, which can be interpreted as a coarse CFO proxy sampled at the byte timescale. Let $f_b\in\mathbb{Z}$ denote the signed \texttt{FREQEST} value associated with byte $b$ of a packet (in device-specific LSB units), and let $y_b\in\{0,1\}^8$ denote the corresponding payload byte. We construct \emph{hamming-weight–conditioned CFO} features by binning these CFO proxies by the byte Hamming weight $w_b=\mathrm{HW}(y_b)=\sum_{k=0}^{7} y_{b,k}\in\{0,\dots,8\}$ and computing class means:
\begin{equation}
\widehat{\Delta f}^{(W)}_{h}
\;=\;
\frac{1}{|\mathcal{B}_h|}
\sum_{b\in\mathcal{B}_h} f_b,
\qquad
\mathcal{B}_h = \{\, b \mid \mathrm{HW}(y_b)=h \,\}.
\label{eq:hweight_cfo}
\end{equation}
Intuitively, conditioning on Hamming weight acts as a \emph{content-aware stratification} of the receiver’s frequency proxy: bytes with different Hamming weights induce different transition densities and symbol trajectories under GFSK (after whitening and modulation shaping), which interact with transmitter/receiver filtering memory and discriminator bias. Therefore, while $f_b$ is a coarse statistic, the vector $\{\widehat{\Delta f}^{(W)}_{h}\}$ captures systematic, hardware-dependent asymmetries that are averaged out by a single global CFO mean. In our implementation, we use a small set of representative bins (e.g., $h\in\{0,2,4,8\}$) through existing USB-exposed fields to avoid protocol changes. 

We condition CFO estimates on the Hamming weight of transmitted bytes, yielding the $5$-tuple feature:
\begin{equation}
\mathbf{f}_{\textsf{HW}}
=
\big[
\widehat{\Delta f}_{\textsf{packet}},\;
\widehat{\Delta f}_{w=0},\;
\widehat{\Delta f}_{w=2},\;
\widehat{\Delta f}_{w=4},\;
\widehat{\Delta f}_{w=8}
\big],
\label{eq:holistic_fingerprint_hw}
\end{equation}
where $\widehat{\Delta f}_{w=k}$ denotes the mean frequency offset accumulated over all bytes of Hamming weight $k$ within the packet. 

\paragraph{Stability and discriminability of features.}
The Ubertooth-based analysis in Fig.~\ref{fig:Uber_cfoTot_violin_persistent} shows that even a coarse, byte-timescale CFO proxy yields a \emph{stable} per-device signature: for most devices, the per-packet CFO distribution is tightly concentrated (narrow violins), while the centers of these distributions differ substantially across devices and tag families (separation on the order of \emph{tens of kHz}), enabling reliable inter-device discrimination despite measurement noise. The Hamming-weight–conditioned features in Fig.~\ref{fig:Uber_weighted_features} further strengthen distinctiveness: within each device, the conditioned CFOs remain consistent across packets (small intra-device spread), yet the \emph{relative pattern} across bins (e.g., $w0$ vs.\ $w2$ vs.\ $w4$ vs.\ $w8$) is markedly device-specific, providing additional dimensions to separate devices whose global CFO partially overlaps. Together, these plots empirically validate that our modulation-aware CFO features are both \emph{persistent} and \emph{discriminating}, making them well-suited as an effective physical-layer fingerprint even under receiver constraints. 

%% file: Sections/Behavior_of_current_tags.tex
\section{Behavior of Commodity Trackers}
\label{sec:tracker_behavior}

\begin{figure*}[t]
    \centering
    \begin{subfigure}[t]{\linewidth}
        \centering
        \includegraphics[width=\linewidth]{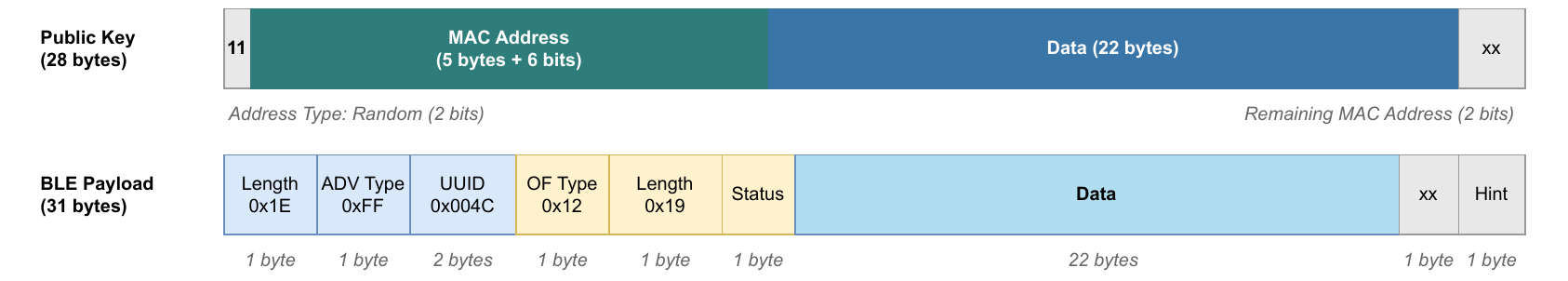}
        \caption{Apple Find My advertisement structure (AirTag-class).}
        \label{fig:pkts-apple}
    \end{subfigure}

    \begin{subfigure}[t]{\linewidth}
        \centering
        \includegraphics[width=\linewidth]{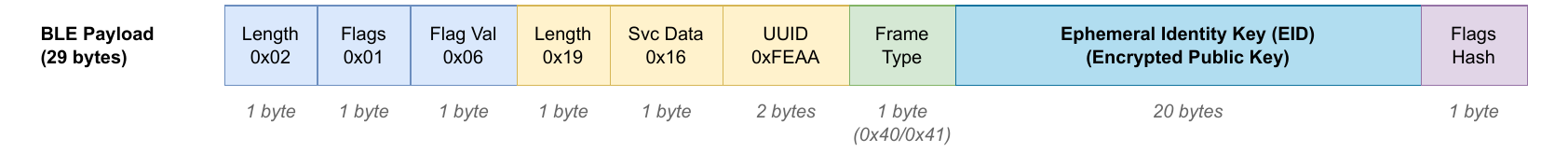}
        \caption{Google Find My Device advertisement structure.}
        \label{fig:pkts-google}
    \end{subfigure}

    \begin{subfigure}[t]{\linewidth}
        \centering
        \includegraphics[width=\linewidth]{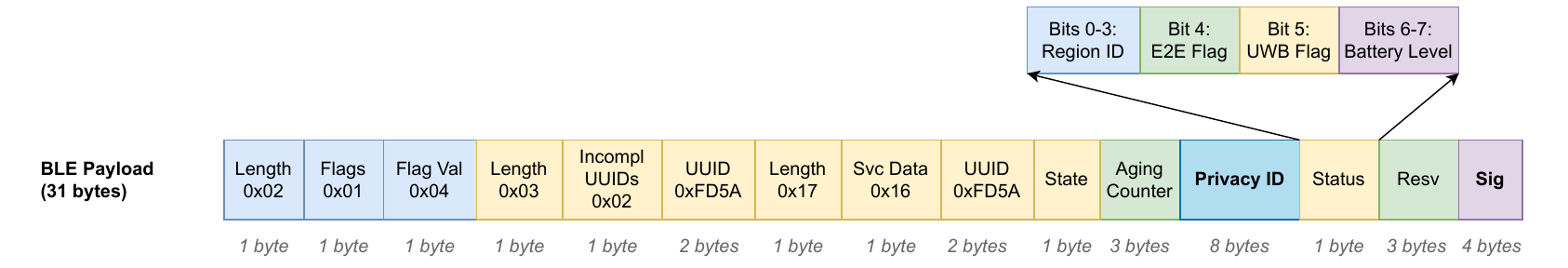}
        \caption{Samsung Find Network advertisement structure (SmartTag2).}
        \label{fig:pkts-samsung}
    \end{subfigure}

    \begin{subfigure}[t]{\linewidth}
        \centering
        \includegraphics[width=\linewidth]{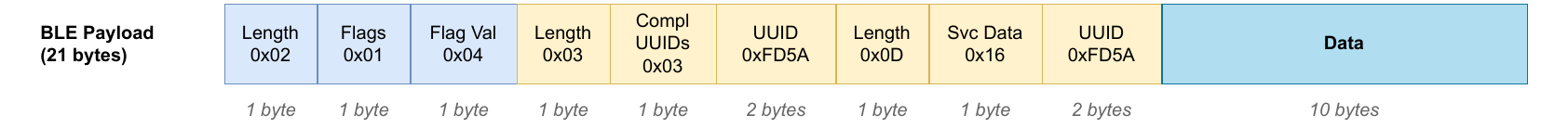}
        \caption{Tile advertisement structure.}
        \label{fig:pkts-tile}
    \end{subfigure}

    \caption{Packet structure for Find My networks across tracker ecosystems.}
    \label{fig:tile-why-track}
    \vspace{-0.8em}
\end{figure*}

We analyze the protocol behavior and privacy mechanisms of widely deployed BLE-based tracking ecosystems: Apple, Samsung, Google, and Tile, focusing on how they randomize link-layer identifiers, rotate protocol-visible cryptographic material, and trigger protections against unwanted tracking. Our analysis is organized along four dimensions: (i) MAC address randomization, (ii) public key / identifier updates, and (iii) stalker detection mechanisms. The corresponding packet layouts that realize these design choices are illustrated in Figure~\ref{fig:tile-why-track} 

\paragraph{System model and terminology.}
Across ecosystems, we distinguish three roles: (i) the \emph{owner device}, which is authenticated and associated with the tracker; (ii) \emph{helper devices}, which periodically scan BLE advertisements and upload encrypted location reports; and (iii) the \emph{tracked device} (the BLE tag). Helper devices do not learn the identity of the tag owner and act solely as relays to the backend infrastructure. 

\subsection{Samsung Find Network}
\label{subsec:samsung}
Samsung’s Find Network~\cite{samsung_usenix} consists of an owner device (a Samsung smartphone), helper devices (nearby Samsung devices), and trackers. Communication between the owner and the tag relies on BLE advertisements and a proprietary GATT profile. SmartTag advertises using Samsung-specific service data formats.

\noindent\textbf{Operational states.}
SmartTag operates in multiple privacy-relevant states: \emph{connected}, \emph{premature lost}, \emph{lost}, and \emph{overmature lost}. After a separation interval, the tag transitions into a ``lost'' reporting regime to enable relay by helper devices; after long separation, it enters an overmature regime where protocol activity is reduced for power efficiency and to support anti-tracking workflows.

\noindent\textbf{Identifier and MAC address rotation.}
SmartTags periodically updates a privacy identifier (PRIVID) and auxiliary fields embedded in BLE service data. The PRIVID remains constant in the lost state, for 24 hours, but notably, SmartTags continue to randomize its BLE MAC address even in lost states, potentially reducing long-term linkability at the link layer if one is only using MAC address.

\subsection{Apple Find My Network}
\label{subsec:apple}
Apple’s Find My network~\cite{airguard, Apple_trackers_aanjhan} consists of Apple owner devices logged into iCloud, helper devices running iOS or macOS, and tracked devices such as AirTags. AirTags rely exclusively on BLE advertisements and do not establish persistent connections with helper devices.

\noindent\textbf{Protocol design.}
AirTags periodically broadcast advertisement payloads containing rotating identifiers derived from ephemeral public keys. The corresponding private keys are stored only on the owner’s device, ensuring helper devices and the backend cannot decrypt location information.

\noindent\textbf{State transitions and safety.}
AirTags transition between near-owner and separated/lost modes based on proximity and time. Advertisement identifiers rotate every 15 minutes in the connected state but stay constant in the lost state; Apple complements cryptographic unlinkability with OS-level anti-stalking defenses (user notifications and audible alerts). Prior work shows that while these mechanisms provide strong unlinkability guarantees, residual tracking risks remain under certain conditions~\cite{airguard,Apple_trackers_aanjhan}.
On the other side, iPhones and Airpods that transmit packets always change their MAC address every 15 minutes. Airpods transmit packets in connected and lost state. Meanwhile, phone transmits when not connected to the network.

\subsection{Google Find My Device Network}
\label{subsec:google}
Google’s Find My Device network~\cite{google_tracker} provisions trackers using Fast Pair and uses public-key material to encrypt location reports relayed by helper devices. BLE advertisements use an ecosystem-specific service format. 

\noindent\textbf{Unwanted Tracking (UT) mode.}
Google introduces a server-controlled Unwanted Tracking (UT) mode that is expected to activate after prolonged separation from the owner and the completion of helper participation. In UT mode, Google trackers do not rotate their Bluetooth MAC address; instead, they retain a fixed address while rotating the cryptographic identifier in the advertising payload approximately every 15 minutes. Moreover, prior work ~\cite{google_tracker} showed that UT activation depends on a server-side timer, which can be reset by submitting a forged “owner present” report, thereby preventing UT mode from triggering. In practice, this design exposes meaningful degrees of freedom for firmware-controlled adversaries, who can balance stealth (e.g., low duty cycle and delayed UT activation) against localization capability.

\subsection{Tile Network}
\label{subsec:tile}
Tile’s ecosystem~\cite{tile_georgia_tech} consists of Tile trackers, owner devices, and helper devices running the Tile application. Unlike Apple, Google, and Samsung, Tile relies on application-level participation rather than OS-level integration, resulting in more limited helper coverage.

\noindent\textbf{Identifier stability and anti-stalking.}
Tile trackers broadcast BLE advertisements containing identifiers that remain stable for extended periods. It does not change its MAC address in connected or lost states, enabling longer-lived linkability. Anti-stalking protections are primarily application-based and require explicit installation and permissions, limiting their effectiveness against passive tracking.

\paragraph{Takeaway.}
Despite different protocol formats and safety triggers, all ecosystems must emit repeated BLE advertisements to remain discoverable by nearby helpers. Ecosystem-specific rotation policies can reduce \emph{identifier-layer} linkability, but they do not eliminate \emph{RF-layer} persistence. These divergences expose distinct attack surfaces, especially for adversaries that exploit fast identifier changes and duty-cycling, which motivates a solution like \textsc{AirCatch} that remains effective under aggressive rotation and stealthy transmission schedules. 